\documentclass[12pt]{article}

\usepackage{epsf}

\newlength{\abstractwidth} 
\renewcommand{\thefootnote}{\fnsymbol{footnote}}
\renewcommand{\thanks}[1]{\footnote{#1}}
\newcommand{\starttext}{
\setcounter{footnote}{0}
\renewcommand{\thefootnote}{\arabic{footnote}}}

\newcommand{\bea}{\begin{eqnarray}}
\newcommand{\eea}{\end{eqnarray}}
\newcommand{\ee}{\end{equation}}
\newcommand{\be}{\begin{equation}}

\def\B{{\cal B}}

\def\G{{\Gamma}}
\def\H{{\cal H}}

\def\Lag{{\cal L}}

\def\N{{\cal N}}
\def\O{{\cal O}}

\def\Re{{\rm Re}}
\def\Im{{\rm Im}}
\def\tr{{\rm tr}}

\def\half{ {1\over 2}}

\def\p{\partial}

\def\ep{\varepsilon}
\def\e{\hat e}

\def\no{\nonumber}

\begin{document}
\starttext
\baselineskip=16pt
\setcounter{footnote}{0}

\begin{flushright}
UCLA/06/TEP/02 \\
22 February 2006
\end{flushright}

\bigskip

\begin{center}
{\Large\bf Ten-dimensional supersymmetric Janus solutions}

\vskip .7in 

{\large  Eric D'Hoker,  John Estes and  Michael Gutperle}

\vskip .2in

 \sl Department of Physics and Astronomy \\
\sl University of California, Los Angeles, CA 90095, USA

\end{center}

\vskip .5in

\begin{abstract}

The reduced field equations and BPS conditions are derived in Type IIB 
supergravity for configurations of the Janus type, characterized 
by an $AdS_4$-slicing of $AdS_5$, and various degrees of internal
symmetry and supersymmetry. A generalization of the Janus solution,
which includes a varying axion along with a varying dilaton, and has $SO(6)$
internal symmetry, but completely broken supersymmetry, is obtained 
analytically in terms of elliptic functions.
% ED revision begin
A two-parameter family of solutions with 4 real
% delete conformal 
% ED revison end
supersymmetries,  $SU(3)$ internal symmetry,
a varying axion along with a varying dilaton, and non-trivial $B_{(2)}$ field,
is derived analytically in terms of genus 3 hyper-elliptic integrals. 
This supersymmetric solution is the 10-dimensional  Type IIB dual
to the $\N=1$ interface super-Yang-Mills theory with $SU(3)$ internal
symmetry previously found in the literature.

\end{abstract}

\vfill\eject

\baselineskip=16pt
\setcounter{equation}{0}
\setcounter{footnote}{0}

\newpage

%%%%%%%%%%%%%%%%%%%%%%%%%%%%%%%%%%%%%%%%%
%%%%%%%%%%%%%%%%%%%%%%%%%%%%%%%%%%%%%%%%%
\section{Introduction}
\setcounter{equation}{0}
%%%%%%%%%%%%%%%%%%%%%%%%%%%%%%%%%%%%%%%%%
%%%%%%%%%%%%%%%%%%%%%%%%%%%%%%%%%%%%%%%%%

The AdS/CFT correspondence  relates string theories on  Anti-de Sitter 
space-times (AdS) to  conformal field theories (CFT) on the boundary of 
the AdS space-time \cite{Maldacena:1997re,Gubser:1998bc,Witten:1998qj}.  
The case which is understood perhaps best relates Type~IIB string theory 
on $AdS_5  \times S_5$ to Yang-Mills theory with $\N=4$ supersymmetry 
and gauge group $SU(N)$. 
(For reviews, see \cite{D'Hoker:2002aw,Aharony:1999ti}.)

\smallskip

The AdS/CFT correspondence is expected to hold as well in situations 
with less or no supersymmetry,  and with space-times which are only 
asymptotically AdS. Thus, a suitable deformation on the gauge theory
side of the correspondence should be dual to an associated deformation 
on the string theory side and vice versa. Many interesting such cases  
are known and have been studied extensively. These include the holographic
representation of renormalization group-flows \cite{Freedman:1999gp}, 
and the solutions of Klebanov-Strassler~\cite{Klebanov:2000hb},
Polchinski-Strassler~\cite{Polchinski:2000uf} and Maldacena-Nunez \cite{Maldacena:2000yy} (first obtained by Chamseddine and Volkov \cite{Chamseddine:1997nm,Chamseddine:1997mc})

\smallskip

In \cite{Bak:2003jk}, a dilatonic deformation of the Type IIB background 
$AdS_5\times S_5$ was found.\footnote{It was named the Janus solution, after
the two-faced Roman god of gates, doors, beginnings, endings and, now, also, 
string dualities.} The Janus solution breaks all supersymmetries, but is 
nevertheless stable against small and a large class of large
perturbations
\cite{Freedman:2003ax,Papadimitriou:2004rz,Celi:2004st}. It can be
viewed as a curved dilatonic domain wall  
\cite{Sonner:2005sj}. 
The holographic dual to the Janus solution is  $\N=4$ super Yang-Mills 
theory in 3+1 space-time dimensions, with a planar 2+1 dimensional  interface,
across which the gauge coupling varies discontinuously. The interface
carries no additional degrees of freedom. Conformal invariance 
in 2+1 dimensions is preserved by the interface at the classical level, 
and holds in conformal perturbation theory to first non-trivial 
order as well  \cite{Clark:2004sb}.

\smallskip

The Janus solution is remarkably simple \footnote{Other dilatonic 
deformations found in the literature are singular and their physical
interpretation remains more obscure \cite{Gubser:1999pk,Bak:2004yf,Kehagias:1999tr}.}. In fact, in this paper, we shall
show that, even when generalized to include a varying axion in addition 
to a varying dilaton,  the Janus solution admits an analytic form in terms 
of elliptic functions. This raises the hope that correlation functions in the 
Janus background may be studied using analytic methods, a topic 
which we plan to address in a later publication.

\smallskip

Furthermore, the remarkable simplicity of the non-supersymmetric Janus 
solution suggests that Janus may possess supersymmetric generalizations 
available in analytic form as well. Clearly, such analytic solutions would
be valuable as starting points for the analytic study of correlators in the 
corresponding backgrounds. (Note that,  on the one hand,  the backgrounds of 
Klebanov-Strassler~\cite{Klebanov:2000hb} or 
Maldacena-Nunez \cite{Maldacena:2000yy} are not asymptotically AdS while,  
on the other hand, the
Polchinski-Strassler~\cite{Polchinski:2000uf} solution is only  known 
approximately, in an expansion in the strength of the fluxes.) 

\smallskip

The fact that interesting supersymmetric generalizations of Janus 
should exist is further suggested by considering its CFT dual, namely 3+1
dimensional super Yang-Mills theory with a 2+1 dimensional planar interface. 
In \cite{Clark:2004sb}, it was found that 2 Poincar\'e supersymmetries can be preserved by adding ``interface  operators" whose support is confined to the interface. 
In the conformal limit, 2 conformal supersymmeries emerge 
as well. The interface operators break the  R-symmetry from  $SO(6)$ down to $SU(3)$. 

\smallskip

In \cite{edjemg}, a complete classification of  supersymmetry restoring 
interface operators on the gauge theory side is given. In particular, it is 
established there that, for interface theories with $SU(3)$ internal symmetry, 
4 is the maximum number of conformal supersymmetries. The corresponding
theory is constructed explicitly, and coincides with the one presented in 
\cite{Clark:2004sb} in terms of $\N=1$ off-shell fields. It is further established 
in \cite{edjemg} that interface theories with extended supersymmetry exist as well. 
One interface theory has 8 conformal supersymmetries and 
$SO(2) \times SU(2)$ internal symmetry, while another has 16 conformal
supersymmetries and $SU(2) \times SU(2)$ internal symmetry.

\smallskip

Further evidence for the existence of 10-dimensional supersymmetric 
generalizations of Janus is provided in \cite{Clark:2005te}, where a 
supersymmetric Janus solution of five dimensional gauged supergravity  
was found (building on previous work on curved domain walls in $AdS_5$
\cite{Behrndt:2002ee,Ceresole:2001wi,Cardoso:2002ec,LopesCardoso:2001rt,Cardoso:2002ff}). 
The starting point of \cite{Clark:2005te} was an $SU(3)$ invariant
gauging of the universal hypermultiplet of  five dimensional $\N=2$
supergravity \cite{Ceresole:2000jd}.  While it is believed that the 
resulting theory is a consistent truncation   
of  the ten-dimensional supergravity, the details of this truncation, and hence of  
any possible lift of the solution to ten dimensions, are unknown.

\smallskip

In the present paper, a family of supersymmetric Janus solutions is derived
directly in ten-dimensional Type IIB supergravity. We focus here on  Janus 
solutions which provide holographic duals to the supersymmetric interface
CFTs of \cite{Clark:2004sb}, with $\N=1$ interface supersymmetry 
and 3 chiral multiplets related by  $SU(3) $ internal symmetry.
Our starting point is the construction of  the most general Ansatz for Type IIB
supergravity fields which preserves $SO(2,3) \times  SU(3) $ symmetry,
and which transforms covariantly under the $SL(2,{\bf R})$ symmetry
of Type IIB supergravity, as well as under the unique $U(1)_\beta \subset SU(4)$
which commutes with $SU(3)$. The reduced field equations are then solved 
subject to the BPS conditions, namely the conditions for the 
vanishing of the supersymmetry variations of the dilatino and gravitino fields.
The resulting family of solutions contains a subset that is of the Janus type,
and the solutions in this subset may be expressed analytically via hyper-elliptic 
integrals of genus 3.
 
\medskip
 
% revision MG + ED %
The Ansatz that will be obtained in this paper for the construction of supersymmetric
Janus solutions is based on an $AdS_{4}$ slicing of $AdS_{5}$, just as the original
non-supersymmetric Janus Ansatz was. If a slicing of $AdS_5$ by 4-dimensional Minkowski
space were used instead, one would recover  an Ansatz used by Romans 
\cite{Romans:1984an} to construct  $SU(3)$-symmetric compactifications of Type IIB supergravity, but without supersymmetry. (In eleven dimensional supergravity, 
the corresponding solutions were constructed  in \cite{Pope:1984jj}.)  
In AdS/CFT such Minkowski-sliced solutions have a natural interpretation in terms of 
RG flows \cite{Freedman:1999gp,DeWolfe:1999cp,Pilch:2000fu,Pilch:2000ue}. 
Some techniques used for the study of the Minkowski slicings can be applied to the 
$AdS_4$ slicings, and the resulting reduced field equations are related.\footnote{We
thank the referee for pointing out this relationship.} 
% revision end %

\smallskip
 
The remainder of the paper is organized as follows. 

\smallskip

In Section 2, the field equations of Type IIB supergravity, as well
as the supersymmetry variations (both for vanishing fermion fields) are summarized.
In Section 3, the original Janus solution is reviewed, 
extended to include a varying axion along with a varying dilaton, 
and expressed analytically in terms of elliptic functions.
In Section 4,  the main results on supersymmetric interface 
CFT are collected.

\smallskip

In Section 5, the most general Ansatz for Type IIB supergravity fields,
subject to $SO(2,3) \times  SU(3) \times U(1)_\beta \times SL(2, {\bf R})$ 
symmetry is constructed. 
The reduced Bianchi identities and field equations for this Ansatz are 
derived in section 6, where it is shown that these equations may also be 
obtained from a reduced action, which is computed, and a vanishing 
Hamiltonian constraint, which amounts to the reduced Wheeler-De Wit equation.
In section~7, the supersymmetry variations for the Ansatz 
are derived. In section 8, it is demonstrated that every supersymmetric
solution with varying axion and varying dilaton is actually the $SL(2, {\bf R})$
image of a ``real" solution with vanishing axion.

\smallskip

In section 9, it is shown that, for supersymmetric solutions, the vanishing 
condition of the Hamiltonian constraint factorizes into a product of two
factors. The vanishing of the first factor leads to a family of degenerate solutions
(for which the second factor vanishes as well), and it is these solutions
which are obtained analytically in terms of genus 3 hyper-elliptic integrals.
In section~10, it is shown that these degenerate solutions are of the janus type,
and thus asymptotically AdS. The equations for the non-degenerate 
solutions are more involved and have not yet been solved analytically.
Numerical evidence suggests that these solutions may not be of the Janus type.
In section 11, the holographic dual CFT is interpreted in terms of 
deformations of the $AdS_5 \times S^5$ background, while in section 12,
some concluding remarks are offered. Finally, a convenient basis
of Dirac matrices is presented in Appendix~A.

\newpage
%%%%%%%%%%%%%%%%%%%%%%%%%%%%%%%%%%%%%%%%%
%%%%%%%%%%%%%%%%%%%%%%%%%%%%%%%%%%%%%%%%%
\section{Type IIB supergravity}
\setcounter{equation}{0}
%%%%%%%%%%%%%%%%%%%%%%%%%%%%%%%%%%%%%%%%%
%%%%%%%%%%%%%%%%%%%%%%%%%%%%%%%%%%%%%%%%%

In this section, we present the field equations, Bianchi identities, and 
supersymmetry variations for Type IIB supergravity, which were
originally derived in \cite{Schwarz:1983qr,Howe:1983sr}. The metric signature 
used here is $(-+ \cdots +)$ in contrast with \cite{Schwarz:1983qr},
where the signature is $(+ - \cdots -)$. We restrict to vanishing fermion fields, 
which will suffice for our analysis.

\smallskip

The bosonic fields of Type IIB supergravity are: the metric $g_{MN}$; 
the axion-dilaton complex scalar $B$ which takes values in the coset 
$SU(1,1)/U(1)$; and the antisymmetric  tensors $B_{(2)}$ 
(which is complex) and  $C_{(4)}$ (which is real). It is standard to introduce composite fields
in terms of which the field equations are expressed simply. They are as 
follows,\footnote{Throughout, we shall pass freely between tensor
and form notations, with a differential form $\omega$ of rank $n$ associated
with tensor  components $ \omega _{M_1 \cdots M_n}$ by the relation
$\omega = {1 \over n!} \omega _{M_1 \cdots M_n} 
dx^{M _1} \wedge \cdots \wedge dx^{M_n}$.}
\bea
P  & =  & f^2 d B, \hskip 1in  f={1\over \sqrt{1-|B|^2}}
\no \\
Q  & = &   f^2 \Im( B d  \bar B). 
\eea
and the field strengths $F_{(3)} = d B_{(2)}$, and
\bea
\label{GF5}
G & = &  f(F_{(3)} - B \bar F_{(3)} )
\no \\
F_{(5)} & = & dC_{(4)} + { i \over 16} \left ( B_{(2)} \wedge \bar F_{(3)} 
- \bar B_{(2)} \wedge  F_{(3)} \right )
\eea
The scalar field $B$ is related to the axion $\chi$ and dilaton $\phi$ fields
by 
\bea
\label{Btau}
B = {1 +i \tau \over 1 - i \tau } \hskip 1in \tau = \tau_1 + i \tau _2 = \chi + i e^{- \phi}
\eea
In terms of the composite fields $P,Q$, and $G$, there are ``Bianchi identities" 
given as follows,
\bea
dP-2i Q\wedge P &=&0
\label{bianchi1} \\
d G - i Q\wedge G +  P\wedge \bar G &=&0
\label{bianchi2} \\
d Q + i P\wedge \bar P& = & 0
\label{bianchi3} \\
d F_{(5)} - i {1\over 8} G\wedge \bar G & = & 0 
\label{bianchi4}
\eea
The field strength $F_{(5)}$ is required to be self-dual,
\bea
\label{SDeq}
 F_{(5) } = * F_{(5) }
\eea
The field equations are given by\footnote{The sign of the term
$GG$ in (\ref{Peq}) has been corrected compared to the
the original equation  in \cite{Schwarz:1983qr}; the need for this correction
was noted independently in \cite{Gauntlett:2005ww,Gran:2005ct,Pilch:2000ue}.}
\bea
0 & = & 
\nabla ^M P_M - 2i Q^M P_M 
+ {1\over 24} G_{MNP }G^{MNP}\label{Peq}
\\
0 & = & 
\nabla ^P  G_{MNP } -i Q^P G_{MNP}
- P^P \bar G_{MNP }
+ {2\over 3} i F_{(5)MNPQR }G^{PQR}
\label{Geq}
\\
0 & = & R_{MN } 
- P_M  \bar P_N  - \bar P_M  P_N 
- {1\over 6} (F_{(5)}^2)_{MN }
\no \\ && \hskip .5in
- {1\over 8} (G_M {} ^{PQ } \bar G_{N PQ }
+ {\bar G_M} {} ^{ PQ } G_{N PQ }) 
+{1\over 48 } g_{MN } G^{PQR } \bar G_{PQR }
\label{Eeq}
\eea
The fermionic fields are the dilatino $\lambda$ and the gravitino $\psi_M$,
both of which are complex Weyl spinors with opposite 10-dimensional
chiralities, given by $\Gamma_{11} \lambda =\lambda$, and $\Gamma_{11}
\psi_M  =-\psi_M$. The supersymmetry variations of the fermions  (still in a
purely  bosonic background) are 
\bea
\delta\lambda
&=& i P_M  \Gamma^M \B^{-1} \ep^* 
-{i\over 24} \Gamma^{MNP }G_{MNP } \ep
\label{susy1} \\
\delta \psi_M 
&=& D _\mu  \ep
+ {i\over 480}F_{(5)NPQRS } 
\Gamma^{NPQRS} \Gamma_M  \ep
+{1\over 96}( \Gamma_M ^{\;\; NPQ }G_{NPQ } 
-9 \Gamma^{NP} G_{MNP }) \B^{-1} \ep^* \qquad
\no
\eea
where ${\cal B}$ is the charge conjugation matrix of the ten dimensional Clifford
algebra.\footnote{It is defined by $\B\B^*=I$ and 
$\B \Gamma ^M \B^{-1} = (\Gamma ^M)^*$; 
see Appendix A for our $\Gamma$-matrix conventions.
Throughout, complex conjugation of functions will be denoted by {\sl bar},
while that of spinors will be denoted by {\sl star}.}

\smallskip

Type IIB supergravity is invariant under $SU(1,1) \sim SL(2,{\bf R})$ symmetry,
which leaves $g_{\mu \nu}$ and $C_{(4)}$ invariant, acts by M\"obius
transformation on the field $\tau$, and linearly on $B_{(2)}$, 
\bea
\label{stransf}
\tau \to {a \tau + b \over c \tau +d} \hskip 1in 
\left ( \matrix{ \Im \, B_{(2)} \cr \Re \, B_{(2)} \cr} \right )
\to 
\left ( \matrix{ a & b  \cr c & d  \cr} \right )
\left ( \matrix{ \Im \, B_{(2)} \cr \Re \, B_{(2)} \cr} \right )
\eea
with $a,b,c,d \in {\bf R}$ and $ad-bc=1$. In this non-linear realization of 
$SL(2,{\bf R})$, the field $B$ takes values in the coset 
$SU(1,1) /U(1) \sim SL(2,{\bf R})/U(1)$,
and the fermions $\lambda$ and $\psi_\mu$ transform linearly under
the isotropy gauge group $U(1)$ with composite gauge field $Q$.

\smallskip

The field equations derive from an action, (we omit the overall prefactor 
$1/2 \kappa _{10}^2 $),
\bea
\label{action1}
S =
\int dx \sqrt{g} \left \{
R - \half { \p_M \tau \p ^M \bar \tau \over (\Im \tau)^2}
- { 1 \over 12} G _{MNP} \bar G ^{MNP}
- 4 | F_{(5)}|^2 \right \}
- i \int C_{(4)} \wedge F_{(3)} \wedge \bar F_{(3)}
\quad
\eea
in the following sense. The field equations are derived by first requiring that $S$
be extremal under {\sl arbitrary variations of the fields  $g_{MN}$, $\tau$, 
$B_{(2)}$ and $C_{(4)}$};  and second by imposing the self-duality condition 
(\ref{SDeq}) on $F_{(5)}$ as a supplementary equation.

\newpage

%%%%%%%%%%%%%%%%%%%%%%%%%%%%%%%%%%%%%%%%%
%%%%%%%%%%%%%%%%%%%%%%%%%%%%%%%%%%%%%%%%%
\section{The generalized non-supersymmetric Janus solution}
\setcounter{equation}{0}
%%%%%%%%%%%%%%%%%%%%%%%%%%%%%%%%%%%%%%%%%
%%%%%%%%%%%%%%%%%%%%%%%%%%%%%%%%%%%%%%%%%

In this section, the original Janus solution of \cite{Bak:2003jk} is reviewed, 
extended to include a varying axion along with a varying dilaton, and expressed analytically in terms of elliptic functions. The Ansatz is required to have
$SO(2,3) \times SO(6)$ symmetry. Since $AdS_4 \times {\bf R} \times S^5$ 
admits no $SO(6)$-invariant  2-forms, the symmetry requires  $B_{(2)}=0$.
The metric $g_{MN}$ and 5-form $F_{(5)}$ are given by an $AdS_4$-slicing 
of $AdS_5$, consistent with $SO(2,3) \times SO(6)$ symmetry,
\bea
\label{janusmetric}
ds^2  
& = & h d\mu^2 + h ds_{AdS_4}^2  + h_1 ds_{S^5}^2
\no \\
F_5 & = & 2 h^{5/2} d\mu \wedge \omega _{AdS_4} + 2 h_1 ^{5/2} \omega _{S^5}
\eea
where $\omega _{AdS_4}$ and $\omega _{S^5}$ are the canonical volume 
forms on the corresponding manifolds. 
The dilaton $\phi$, axion $\chi$ and the functions $h, h_1$ depend on $\mu$ only. 
Remarkably, the original Janus solution was found by setting $h_1=1$, thereby
leaving the $S^5$ metric unchanged. Numerical evidence suggests that solutions with varying $h_1$ always have singularities. In section \S 3.3, we shall present  arguments, based on the AdS/CFT correspondence, that such breathing modes 
which vary $h_1$, must be absent lest the corresponding solution become singular.

\smallskip

The reduced field equations (for $h_1=1$) may be expressed in terms of $\tau$ 
and $h$,
\bea
{\tau '' \over \tau '} + { i \over \tau_2} \tau ' +  {3 \over 2} \, {h' \over h} & = & 0
\label{taueq} \\
4 (h')^2 - 4 h h'' + 8 h^3 & = & h^2 \, {|\tau'|^2 \over  \tau_2^2}  
\label{eqh1} \\
12 h^2 + (h')^2 + 2 h h'' - 16 h^3 & = & 0
\label{eqh2}
\eea
Since $h$ is real, the imaginary part of (\ref{taueq}) is independent
of $h$ and may be integrated to give 
\bea
|\tau - p |^2 = r^2 \hskip 1.4in p,r \in {\bf R}
\eea
This means that as $\mu$ varies, $\tau$ evolves along a segment of a 
geodesic in the upper half plane equipped with the $SL(2, {\bf R})$-invariant 
Poincar\'e metric\footnote{Its geodesics are the half circles with  arbitrary center 
$p$ on the real axis and arbitrary radius $r$.}  
$|d\tau|^2/\tau_2^2$. Integrating also the real part 
of (\ref{taueq}), we find that $|\tau' |^2 /\tau_2^2 = c_0^2 / h^3$.
for some  constant $c_0 \in {\bf R}$. This relation gives the velocity of
$\tau$ along the geodesic as a function of $h$. Finally, using this result
in (\ref{eqh1}) leads to an equation that is consistent with (\ref{eqh2})
and has a first integral given by,
\bea
\label{januseq}
(h')^2 = 4 h^3 - 4h^2 + {c_0^2 \over 6 h}
\eea
which is the same equation as the first integral for the original
Janus solution \cite{Bak:2003jk}.

\smallskip

The interpretation of the above results is as follows. In the original  Janus 
solution, the axion field vanished, and the dilaton evolution spanned a rather 
special geodesic in the upper half plane: a vertical line segment with $\tau_1=0$.
Since the action of  $SL(2,{\bf R})$ on the upper half plane is transitive
on points as well as on connected geodesic segments of equal length, 
solutions with varying axion may be obtained as 
$SL(2, {\bf R})$ images of solutions with vanishing axion. The remarkable 
result obtained above is that {\sl all solutions} with varying axion may be 
obtained as $SL(2, {\bf R})$ images of solutions with vanishing axion.

\begin{figure}[tbph]
\begin{center}
\epsfxsize=4.5in
\epsfysize=4.1in
\epsffile{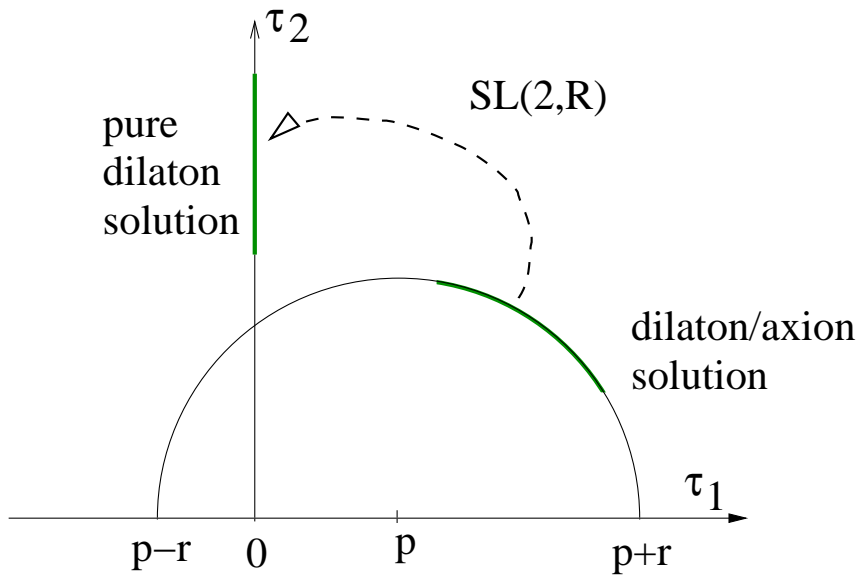}
\label{figure1}
\vskip -.9in
\caption{Mapping geodesic segments under $SL(2,{\bf R})$ in the 
dilaton/axion upper half  plane.}
\end{center}
\vskip -.3in
\end{figure}

\subsection{Analytical solution in terms of elliptic functions}

In \cite{Bak:2003jk}, the quadrature of (\ref{januseq}) was obtained 
numerically. Actually, the general solution may be expressed 
analytically in terms of elliptic functions. To do so, we consider a
coordinate independent object, namely the 1-form,
\bea
\nu = {d \mu \over \sqrt{h}}  = { dh \over \sqrt{ 4 h^4 - 4 h^3 + c_0^2/6}}
\eea
The 1-form $\nu$ is proportional to the holomorphic  Abelian  differential on a torus. 
We represent the constant $c_0$ by $c_0^2 = 24 \gamma _0^3(1-\gamma _0)$, 
and parametrize the Abelian differential $\nu$ and the function $h$ in terms of $\gamma _0$ and  the Weierstrass $\wp$-function, expressed in terms of the canonical coordinate $z$ of the torus. We find the following expressions, 
\bea
\nu = {dz \over \sqrt{\gamma _0}}
\hskip 1in
h(\mu) = 
\gamma _0 + {\gamma _0 (3-4\gamma _0) \over \wp (z) + 2\gamma _0 -1}
\eea
Here, the Weierstrass $\wp$-function has been normalized to standard form,
$(\p_z \wp )^2 = 4 \wp ^3 - 16\gamma _0 (1- \gamma _0) \wp - 4(1-\gamma _0)$. The discriminant of the corresponding curve 
$\Delta = 64 c_0^4 (32 c_0^2 - 81)/27$ vanishes at $c_0^2=0$ and at the critical 
value $c_0^2 = 81/32$, identified in \cite{Bak:2003jk} as the value where the 
range of the dilaton begins to diverge. 
The dilaton axion equation reduces to
\bea
| d \tau |/\tau_2 = c_0 \, {\nu \over h} 
\eea
which may be integrated by standard elliptic function methods.

\subsection{Structure of the AdS/CFT dual}
\label{secadscftdual}

The Janus solution can be viewed as a dilatonic domain wall in which the dilaton 
varies  with the coordinate $\mu$, which parameterizes the $AdS_4$ slicing 
of $AdS_5$. It follows from the dilatino supersymmetry variation
(\ref{susy1}), that no supersymmetries are preserved for the Janus
solution with a varying dilaton and vanishing $B_{(2)}$.
Nevertheless, in \cite{Freedman:2003ax,Papadimitriou:2004rz,Celi:2004st}  convincing arguments were presented that the Janus 
solution is stable against all small and a certain class of large
perturbations.  

\begin{figure}[tbph]
\begin{center}
\epsfxsize=5.5in
\epsfysize=2in
\epsffile{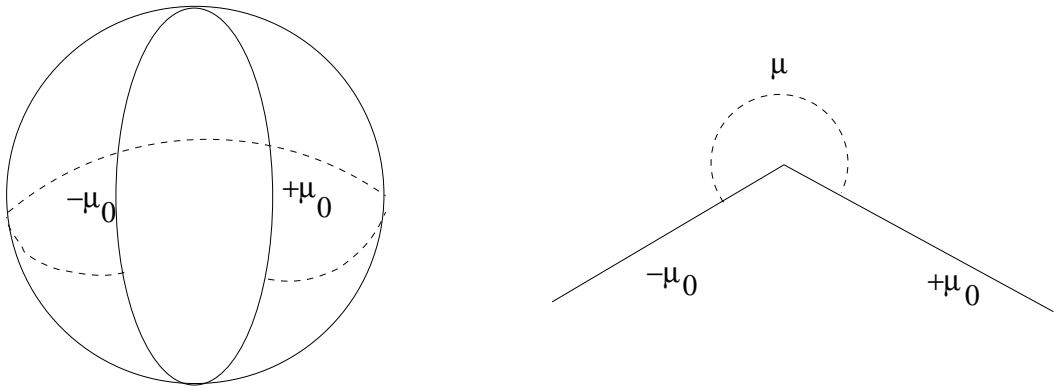}
\label{figure2}
\caption{Sketch of the boundary geometry of the Janus solution in global and
  Poincare coordinates for the $AdS_4$ slices.}
\end{center}
\end{figure}

The angular coordinate $\mu$  covers a range $\mu\in[-\mu_0,\mu_0]$ where $\mu_0> \pi/2$.  The structure of the boundary of this space can be analyzed 
using  global coordinates for the $AdS_4$ slices.  Near $\mu=\pm \mu_0$ 
the noncompact part of the metric has the following asymptotic behavior,
\be
ds^2\sim {1\over (\mu\mp \mu_0)^2 \cos^2\lambda }\big( \cos^2\lambda
d\mu^2 - dt^2 + d\lambda^2 + \sin^2\lambda d\Omega^2_{S_2}\big) 
\ee
with $0\leq \lambda <{\pi\over 2}$. The constant time section of the boundary 
is a non-singular geometry, consisting of two halves of $S^3$ at $\mu =\pm \mu_0$, joined  at the pole of $S_3$  where $\lambda=\pi/2$. In Poincar\'e coordinates 
for the $AdS_4$  slices, the spatial section of the boundary consists of two three dimensional  half planes joined by a two dimensional interface.
The dilaton varies continuously with $\mu$ and takes two different
constant values at the boundary,
\be
\lim_{\mu\to \pm \mu_0}\phi(\mu) = \phi^{(0)}_{\pm} + \phi^{(1)}_\pm
(\mu\mp \mu_0)^4 + O [(\mu\mp \mu_0)^8]
\ee 
The holographic dual gauge theory CFT of the Janus solution was proposed in
\cite{Bak:2003jk} and analyzed in detail in \cite{Clark:2004sb}.
The CFT dual is a planar interface theory. The action on both sides of the 
interface is the standard $\N=4$ SYM action but the coupling constant varies
discontinuously across the interface.  The symmetry $SO(2,3)$ of the Janus 
solution maps to the conformal symmetry of a planar interface on the CFT side.
This symmetry is manifest at the classical level, but was also shown to
persist at the first non-trivial quantum level \cite{Clark:2004sb}.
The $SO(6)$ symmetry of the Janus solution maps to an (accidental)
internal symmetry on the CFT side. 
Note that, in contrast to the defect conformal field theories examined earlier 
in the context of AdS/CFT \cite{Karch:2000gx,DeWolfe:2001pq,Aharony:2003qf,Erdmenger:2002ex,Yamaguchi:2003ay},
the CFT dual to the Janus solution is characterized by an interface that carries 
no degrees of freedom in addition to the ones inherited from the bulk $\N=4$,
whence the name ``interface", as opposed to ``defect".

\subsection{The absence of breathing modes}

The AdS/CFT dictionary relates the  breathing mode of $S^5$, described 
by the function $h_1$ in the Janus metric of (\ref{janusmetric}), to a dimension 
8 operator $\O^{(20)}_{k=0}=\tr(F_+^2 F_-^2)$ (in the notation of 
\cite{D'Hoker:2002aw}). 
The argument below will show that the breathing mode cannot be excited in 
the Janus solution, and thus the function $h_1$ must be a constant.
On the one hand, a non-vanishing source for the operator $\O^{(20)}_{k=0}$ 
would correspond to a behavior $(\mu-\mu_0)^{-4}$ near the AdS boundary, as
$\mu\to \mu_0$. Such a behavior would lead to a singular 10-dimensional 
metric. Since the Janus solution is regular, this source must be absent.
On the other hand, a non-vanishing expectation value would correspond to a 
behavior $(\mu-\mu_0)^8$ near the boundary.  A power series expansion near the boundary of AdS for the Janus solution, and allowing for the presence of a 
breathing mode, reveals that  such a term is forced to vanish. By standard 
AdS/CFT arguments the breathing mode is therefore exactly zero.

\newpage

%%%%%%%%%%%%%%%%%%%%%%%%%%%%%%%%%%%%%%%%%
%%%%%%%%%%%%%%%%%%%%%%%%%%%%%%%%%%%%%%%%%
\section{Supersymmetric interface CFT}
\setcounter{equation}{0}
%%%%%%%%%%%%%%%%%%%%%%%%%%%%%%%%%%%%%%%%%
%%%%%%%%%%%%%%%%%%%%%%%%%%%%%%%%%%%%%%%%%

In \cite{Clark:2004sb}, it was shown that preserving supersymmetry in a planar 
interface Yang-Mills theory necessarily leads to the breaking of the internal 
$SO(6)$. In turn, arguments are presented in \cite{Clark:2004sb} that 2 
Poincar\'e supercharges may be preserved upon reducing $SO(6)$ to $SU(3)$, and 
including certain ``interface counterterms" which are built out of the fields of 
the bulk $\N=4$  super Yang-Mills theory.

\smallskip

Specifically, in \cite{Clark:2004sb}, the Yang-Mills coupling $g(x^\pi)$ is 
assumed to be a function of the coordinate $x^\pi$ 
%ED revision begin
(the CFT side coordinate corresponding to the coordinate $\mu$ of the AdS side)
%ED revision end
and to vary across the interface located at $x^\pi=0$. The Lagrangian is formulated 
with $\N=1$ auxiliary fields  for a single chiral multiplet  and a single gauge multipet, 
\bea
\Lag _{{\rm chiral}} & = & 
- \p_\mu \bar \phi \, \p^\mu \phi  - { i \over 2} \bar \psi \gamma ^\mu \p_\mu \psi
+ \bar F\, F + \left ( W' F - {i \over 4} W'' \bar \psi (1 + \gamma ^5) \psi + {\rm c.c.} \right )
\no \\
\Lag_{{\rm gauge}} & = &
- {1 \over 4 g^2} F^a _{\mu \nu} F^{a\mu\nu} 
-{ i \over 2 g^2} \bar \lambda ^a \gamma ^\mu D_\mu \lambda ^a 
+ {1 \over 2 g^2} D^a D^a
\eea
Here, $\psi$ and $\lambda ^a$ are Majorana spinors, $\phi$ is a complex 
scalar, $F$ and $D^a$ are auxiliary fields, and $W$ is the superpotential 
of the chiral multiplet. 
In $\Lag_{{\rm chiral}}$, all dependence on the coupling $g$ is contained in $W$. 
Upon adding to $\Lag_{{\rm chiral}}$ and $\Lag _{{\rm gauge}}$ 
the following ``interface counterterms", 
\bea
\delta \Lag _{{\rm chiral}} & = & 
i \p_\pi g \left ( {\p W \over \p g} - {\p \bar W \over \p g} \right )
\no \\
\delta  \Lag_{{\rm gauge}}  & = & 
- \p_\pi \left ( g^{-2} \right ) \left ( 
{i \over 4} \bar \ep \gamma ^\pi \gamma ^{\mu \nu} \lambda ^a F_{\mu \nu} 
+ \half \bar \ep \gamma ^\pi \gamma ^5 \lambda ^a D^a \right )
\eea
the combined Lagrangians are invariant under the supersymmetry 
generated by spinors $\ep$ satisfying the interface projection relation,
$1/2 ( 1 + i \gamma ^5 \gamma ^\pi ) \ep = \ep$. For the $\N=4$ theory, 
we have $W \sim \epsilon ^{ijk} \Phi ^i \Phi ^j \Phi ^k$,
where $\Phi^k$ are 3 complex fields, $\Phi ^k = \phi ^{2k-1} + i \phi ^{2k}$. 
This theory has $SU(3)$ internal symmetry, which rotates the three chiral 
multiplets into one another.

\smallskip

Surprisingly, the derivation of the existence of the $\N=1$ interface supersymmetry
presented in \cite{Clark:2004sb} seems to depend on the precise normalizations
of the chiral and gauge multiplet Lagrangians: canonical for the chiral multiplet,
but with the gauge coupling factored out for the gauge multiplet.

\smallskip

In a companion paper \cite{edjemg}, the existence of supersymmetry in
the $\N=4$ super Yang-Mills theory with an interface and with ``interface  counterterms" is solved in generality. We confirm that
the model of \cite{Clark:2004sb} indeed possesses $\N=1$ interface 
supersymmetry  and $SU(3)$ global symmetry, independently of any 
normalization issues. Further models with supersymmetry, including 
one with $\N=4$ interface supersymmetry and $SO(4)$ internal 
symmetry, are also discovered in  \cite{edjemg} along with a complete
classification of all possible supersymmetries.

\smallskip

%revision MG
It may be helpful to clarify  the counting of the number of preserved supersymmetries. 
Since the interface field theory has only 2+1 dimensional Poincar\'e invariance, 
the counting of supersymmetries is conveniently carried out from a three 
dimensional point of view.
An $\N =1$ interface supersymmetry corresponds to 2 real Poincar\'e supercharges. 
When, in addition, the interface field theory is  conformal invariant, 
the number of supercharges is double the number of Poincar\'e supercharges.
Thus, the $\N=1$ interface CFT has altogether 4 real supercharges. 
In the dual supergravity this means that we look for, and find,  a two-parameter 
family of solutions  which preserve 4 real supersymmetries.
%end revision

\newpage

%%%%%%%%%%%%%%%%%%%%%%%%%%%%%%%%%%%%%%%%%%
%%%%%%%%%%%%%%%%%%%%%%%%%%%%%%%%%%%%%%%%%%
\section{The ten-dimensional Janus Ansatz}
\setcounter{equation}{0}
%%%%%%%%%%%%%%%%%%%%%%%%%%%%%%%%%%%%%%%%%%
%%%%%%%%%%%%%%%%%%%%%%%%%%%%%%%%%%%%%%%%%%

In this section, we shall construct the most general Ansatz for Type IIB
supergravity fields, consistent with the symmetries of the expected CFT 
dual theory with $\N=1$ interface supersymmetry and global $SU(3)$ symmetry
relating the 3 chiral multiplets inherited from the parent $\N=4$ theory.

\subsection{Symmetries of the Ansatz}

For given gauge and interface couplings, the dual CFT has $SO(2,3)$ 
conformal and $SU(3)$ internal symmetry, along with 2 Poincar\'e and 2 conformal
supersymmetries. Therefore, on the AdS side, we shall seek an Ansatz
for the supersymmetric generalization of the Janus solution
which is invariant under the following bosonic symmetries,
\bea
SO(2,3) \times SU(3) 
\eea
The supersymmetries will be achieved  later by enforcing the BPS conditions.
In analogy with Janus, the solution is expected to depend continuously on at least one parameter (for Janus, this is the constant $c_0$) and include the undeformed $AdS_5 \times S^5$ as a limiting case (for Janus, $c_0 \to 0$). Therefore, 
the topology of the internal space is expected to remain the same and equal to the topology of $S^5$. 

\smallskip

The interface CFT naturally consists of a family of theories. This is familiar 
from the parent $\N=4$ theories, which are labeled by the gauge coupling 
$g$ and the instanton angle $\theta$, and are mapped into one another by the standard action of $SL(2,{\bf Z})$ on $g$ and $\theta$. 
Montonen-Olive duality 
states that two theories related by  $SL(2, {\bf Z})$ are physically the same. 
On the AdS side, $SL(2, {\bf Z})$ degenerates to $SL(2, {\bf R})$, which 
constitutes a symmetry of Type IIB supergravity. 
Therefore, on the AdS side, we shall seek an Ansatz which forms a family on which 
$SL(2, {\bf R})$ acts consistently as well. For example, our previous generalization
of Janus, which includes the axion and the dilaton, is such a family of
Ans\"atze, while the original Janus Ansatz clearly is not, since $SL(2, {\bf R})$
does not act consistently on it.

\smallskip

The interface CFT is also naturally a family of theories in terms of its interface couplings. Indeed, the interface couplings of the CFT dual theory of 
\cite{Clark:2004sb} and \cite{edjemg} are mapped into one another by $SU(4)$. Theories with different interface couplings related in this way by $SU(4)$ are physically the same. On the AdS side, not all of these $SU(4)$ transformations 
can be  implemented in a useful way. Once an Ansatz has been forced to be
invariant under $SU(3)$, the embedding of $SU(3)$ in $SU(4)$ is fixed,
and the only useful transformations of $SU(4)$ that can be implemented
on the $SU(3)$-invariant Ansatz are those that commute with $SU(3)$.
This is a single generator, spanning a group  $U(1)_\beta$, the notation of 
which  will be motivated later on. To summarize, we shall seek an Ansatz 
for the  supersymmetric generalization of the Janus solution which
is invariant under the group
\bea
SO(2,3) \times SU(3) \times U(1)_\beta \times  SL(2, {\bf R}) 
\eea
in the sense described above.

\smallskip

The presence of the $SO(2,3)$ factor requires the Ansatz to be an $AdS_4$ 
slicing of $AdS_5$, as was already the case for the original Janus. 
The group $SU(3) \times U(1)_\beta$ must be realized as an isometry of 
the 5-dimensional internal space. The only 4-dimensional manifold with 
$SU(3)$ isometry is  $CP_2 = SU(3) / S(U(2) \times U(1)$. The product 
space $CP_2 \times S^1$ thus has the correct isometry $SU(3) \times U(1)_\beta$, 
but not the correct topology of $S^5$. The  topology of $S^5$ is easily 
recovered by recalling  that $S^5$ is  the total space of a $S^1$ bundle over 
$CP_2$ \cite{Romans:1984an,Pope:1984jj,Liu:2000gk,Pilch:2000fu}. 
A natural candidate internal space for our Ansatz is 
$CP_2 \times _q S^1$, where $\times _q$ produces the $S^1$ bundle 
over $CP_2$ with integer first  Chern class $q \in {\bf Z}$. The requirement
that this space have the topology of $S^5$ reduces this choice to $q=\pm 1$,
so we may simply set $q=1$. (Note that another candidate internal space
$SU(3)/SO(3)$ has the topology of $S^5$, but its isometry does not
include the $U(1)_\beta$ factor.) 

\smallskip

Based on these symmetry considerations, 
we conclude that the Ansatz must be constructed on the space
\bea
{\bf R} \times AdS_4 \times CP_2 \times _1 S^1
\eea
where $CP_2 \times _1 S^1$ is the sphere $S^5$, 
deformed while preserving $SU(3)\times U(1)_\beta$ isometry.

\subsection{Invariant Metrics and Frames on $CP_2$}

In this section, we summarize important properties of $CP_2$ and its
$S^1$ fiber bundle $S^5$, and derive all the invariants needed for the 
construction of the invariant Ansatz. Recall that both spaces may be 
viewed as symmetric spaces via the following cosets $S^5 = SO(6)/SO(5)$ 
and $CP_2 = SU(3)/S(U(2)\times U(1))$. But, $S^5$ may also be viewed 
as a {\sl non-symmetric} homogeneous space via the coset $S^5 = SU(3)/SU(2)$,
from which it is clear that $S^5$ is the total space of a $S^1=U(1)$ bundle 
over $CP_2$. 

\smallskip

The space $CP_2$ may be parametrized locally by two complex coordinates
$\zeta _1, \zeta _2 \in {\bf C}$ or, equivalently, by four real angles 
$\alpha, \theta, \phi, \psi$, related to one another by
\bea
\zeta_1 & = & \tan(\alpha) \cos(\theta/2)e^{{i\over 2}(\psi+\phi)}
\no \\
\zeta_2 & = & \tan(\alpha) \sin(\theta/2)e^{{i\over 2}(\psi - \phi)}
\eea
The Fubini-Study metric is given by
\bea
ds^2 _{CP_2} = g_{i\bar j} d\zeta^i\otimes d \bar \zeta^{\bar j} 
\hskip 1in 
g_{i\bar j}=\partial_i \partial_{\bar j} \ln(1+|\zeta_1|^2+|\zeta_2|^2)
\eea
It is useful to work with an orthonormal frame $\e ^a, \, a=6,7,8,9$ on $CP_2$
which may be expressed in terms of the angular coordinates,\footnote{We choose
the labels 6, 7, 8, 9 for internal labels, as it is in this manner that
the indices will be embedded in 10 dimensional space-time.}
\bea
\label{CP2frame}
\e ^6 &=& d\alpha 
\no \\ 
\e ^7 &=& {1 \over 4} \sin(2\alpha) \sigma_3 
\hskip 1.12in
\sigma_3 = d\psi+\cos(\theta) d\phi
\no \\
\e ^8 & = &  \half \sin(\alpha) \sigma_1
\hskip 1.2in
\sigma_1 = -\sin(\psi) d\theta +\cos(\psi) \sin(\theta)d\phi
\no \\
\e ^9 & = & \half \sin(\alpha) \sigma_2
\hskip 1.2in
\sigma_2 = \cos(\psi) d\theta +\sin(\psi) \sin(\theta)d\phi
\eea
The 1-forms $\sigma ^1, \sigma ^2, \sigma ^3$ span the frame
of $S^3$. We shall also make use of a basis in which the complex structure
of $CP_2$ is manifest, and introduce a complex splitting of the frames,
\bea
\label{complexframe}
\e^{z_1} = \e^6+i \e^7 & \quad & \e^{z_2}= \e^8+i \e^9
\no \\
\e^{\bar z_1} = \e^6-i \e^7 & \quad & \e^{\bar z_2}= \e^8-i \e^9
\eea
The metric  may be expressed in terms of  the above coordinates and frames by,
\bea
ds^2 _{CP_2} & = &
d\alpha^2 +{1\over 4}\sin^2(\alpha) \left (\sigma_1^2+\sigma_2^2 +\cos^2(\alpha)\sigma_3^2 \right )
\no \\
& = &
\sum_{a=6}^9  \e^a\otimes \e^a 
= \e^{z_1} \otimes \e^{\bar z_1} + \e^{z_2}\otimes \e^{\bar z_2}
\eea
We shall also need the torsion-free connection associated with 
this frame, satisfying 
\bea
d \e^a + \hat \omega ^a {}_b \wedge \e ^b=0
\eea
These connection components are given by,
\bea
\label{CP2connection}
\hat \omega ^6 {}_7 = - \half \cos (2 \alpha) \, \sigma _3
& \hskip 1in &
\hat \omega ^6 {}_8 = + \hat \omega ^7 {}_9 = - \half \cos ( \alpha) \, \sigma _1 
\no \\
\hat \omega ^8 {}_9 = \left  ( 1 -  \half \cos^2 ( \alpha) \right ) \, \sigma _3
& \hskip 1in &
\hat \omega ^6 {}_9 = - \hat \omega ^7 {}_8 = - \half \cos ( \alpha) \, \sigma _2 ~
\eea

\subsection{Invariant 2-forms on $CP_2$ and $S^5$}

In this subsection, we shall obtain the most general 2-forms on $S^5$,
invariant under $SU(3)$, and use them to build an 
$SU(3) \times U(1)_\beta$-invariant Ansatz for the antisymmetric tensor field $B_{(2)}$ in the  subsequent section.
It is well-know that the cohomology of $CP_2$ is generated by the K\"ahler 
form $K$, which derives from a $U(1)$-connection $A_1$ by $K=dA_1$.
In terms of the coordinates of $CP_2$ introduced earlier, these forms are
given by,
\bea
\label{K&A1}
K & = & i g_{i \bar j} \, d \zeta ^i \wedge d \bar \zeta ^{\bar j}
\hskip .7in 
=
\half \sin (2 \alpha) d\alpha \wedge \sigma _3 
+ \half \sin^2 (\alpha) \sigma _1 \wedge \sigma _2
\no \\
A_1 & = &  - {i\over 2} \, {\bar\zeta^i d\zeta^i -\zeta^i d\bar \zeta^i \over  1+|\zeta_1|^2+|\zeta_2|^2 }
\hskip .2in 
=  {1\over 2}\sin^2(\alpha) \sigma_3
\eea
In terms of the frames $\e^a$, we have 
\bea
\label{Kahler}
K =  2\, \e^6 \wedge \e^7 + 2\, \e^8 \wedge \e^9
= 
i \e^{z_1}\wedge \e^{\bar z_1}+ i \e^{z_2}\wedge \e^{\bar z_2}
\eea
The K\"ahler form is simply related to the volume form as follows, $K^2 = 8 \e^6 \wedge \e^7 \wedge \e^8 \wedge \e^9$. 

\smallskip

The five sphere $S^5$  is constructed as a $S^1$ fibration over $CP_2$. 
Introducing the $S^1$ coordinate $\beta$, the isometry group $U(1)_\beta$
acts on $S^1$ by shifts of $\beta$. The round metric on $S^5$ 
is then given in terms of this fibration by,
\be
ds^2 _{S^5}= (d\beta + A_1)^2 +ds_{CP_2}^2
\ee
where $A_1$ is the K\"ahler one form defined in (\ref{K&A1}).
We introduce a fifth frame component
\be
\e^5 =  d\beta+A_1 \label{fibcp2}
\ee
Under the action of $SU(3)$, the connection $A_1$ shifts by a non-trivial 
exact form, which may be compensated for by the opposite shift in $\beta$,
so that the combination  $d\beta +  A_1$ is invariant under $SU(3)$.
It is of course also invariant under constant shifts of $\beta $, forming 
the group $U(1)_\beta$. The frame $\e^5, \e^6, \e^7, \e^8, \e^9$ is an orthonormal
frame for $S^5$.

\smallskip

Clearly, the K\"ahler form $K$ on $CP_2$ is invariant under $SU(3)$ and will 
be a candidate for an invariant 2-form in the Ansatz for the antisymmetric 
tensor field $B_{(2)}$. To ensure a consistent Ansatz, however, we shall need 
the {\sl most general 2-form invariant under $SU(3)$ on the deformed sphere $S^5$}, and this requires an exhaustive study of all $SU(3) $-invariant 2-forms 
on $S^5$. In this task, we are helped by two theorems 
(see for example the corollaries 1.18 and 1.14 of  \cite{DFN}, as well as
\cite{D'Hoker:1995it}), valid for compact, connected $G$ and $H$,

\smallskip

{\bf Theorem I} \ 
The ring of $G$-invariant $n$-forms on a homogeneous space $G/H$ is 
obtained from a basis of left-invariant 1-forms $\theta ^a, \, a = 1, \cdots, {\rm dim} G$ on $G$, by constant tensors $\omega _{a_1 a_2 \cdots a_n}$, which vanish whenever $a _i \in \H$, and are invariant under $\H$, by the expression,
\bea 
\omega ^{(n)} = \omega _{a_1 a_2 \cdots a_n} \theta ^{a_1} \wedge \cdots 
\wedge \theta ^{a_n}
\eea
\indent {\bf Theorem II} \
On a symmetric space $G/H$, every $G$-invariant form is  closed.

\medskip

Given that $S^5 = SO(6)/SO(5)$ is a symmetric space coset, and that 
$H^2(S^5, {\bf R})=0$, there are no $SO(6)$-invariant 2-forms on $S^5$, 
a fact that was used to set $B_{(2)}=0$ in the original Janus solutions. 
Given, that the coset $CP_2= SU(3)/S(U(2)\times U(1))$ is a symmetric space, 
and that $H^2(CP_2, {\bf R})$ is generated by a single element,
namely the K\"ahler form $K$, we conclude that $K$ is the unique
$SU(3)$-invariant 2-form on $CP_2$. These facts are standard.

\smallskip

Finally, we wish to obtain all $SU(3)$-invariant 2-forms on $S^5$.
To this end, we express $S^5$ as the coset $S^5 = SU(3)/SU(2)$,
where $SU(2)$ is embedded in $SU(3)$ in the 2-dimensional representation.
This coset is {\sl not a symmetric space}, so that $SU(3)$-invariant forms
need not be closed. It is now straightforward to obtain all such
invariant forms, using Theorem I. We need all $SU(2)$-invariant tensors
on the left-invariant 1-forms $\theta ^a$ on $SU(3)$, which vanish on $SU(2)$.
These 1-forms are precisely $\e^{z_1}, \e^{z_2}, \e^{\bar z_1} , \e^{\bar z_2},
\e ^5$ constructed earlier. Clearly, we recover the K\"ahler form of (\ref{Kahler}) 
following the construction of Theorem I in this manner. There is also the 
following 2-form (and its complex conjugate),
\bea
\e^{z_1} \wedge  \e^{z_2} = \half \sum _{i,j=1,2} \ep _{ij} \e^{z_i} \wedge \e^{z_j}
\eea
which is not invariant under the action of $SU(3)$ isometries on $CP_2$,
because the form is not invariant under the $U(1)$ factor of the isotropy
group of $CP_2$. It can be made into a well-defined $SU(3)$-invariant  form 
$\hat A_2$ on $S^5$ by compensating for the phase factor, 
\bea
\hat A_2  & \equiv  &
  \e^{z_1}  \wedge \e ^{z_2} \, e^{3 i \beta } 
=
 { i \, d \zeta ^1 \wedge d \zeta ^2 \, e^{3 i \beta} \over ( 1+ |\zeta _1|^2 + |\zeta _2|^2)^{3/2}}
\eea
Some further useful properties involving $\hat A_2$ are collected below.
\bea
d \hat A_2 & = &  3 i \, (d \beta + A_1) \wedge \hat A_2
= 3 i  \, \e^5 \wedge \hat A_2
\no \\
\hat A_2 \wedge {\bar {\hat A_2} }
& = &  \half K^2 = 4 \e^6 \wedge \e^7 \wedge \e^8 \wedge \e^9
\eea
In particular, the formula for the differential shows that $\hat A_2$
is indeed the solution to an $SU(3)$-invariant equation 
$(d - 3 i \hat e^5) \hat A_2=0$, which is consistent with the fact that 
$\hat A_2$ itself is invariant. By contrast, the form $\e^{z_1} \wedge  \e^{z_2}$
satisfies a differential equation $(d - 3 i A_1) (\e^{z_1} \wedge  \e^{z_2})=0$
which is not invariant. Under $U(1)_\beta$, the form $\hat A_2$
transforms with a constant phase factor.

\subsection{The Ansatz for the metric}

The Ansatz for the metric follows from the symmetry considerations above,
\bea
ds^2
= f_4^2  \big( d\mu ^2+ds_{AdS_4}^2\big) + f_1^2 (d\beta+  A _1 )^2
+ f_2^2 ds_{CP_2}^2 \label{tendmet}
\eea 
Invariance of the metric under $SO(2,3) \times SU(3) \times U(1)_\beta$ 
requires that the  functions $f_1,f_2$ and $f_4$ depend only on $\mu$.
According to the transformation rules of Type IIB supergravity, the metric,
in the Einstein frame,
must be invariant under $SL(2, {\bf R})$, which requires that the functions
$f_1,f_2$, and $f_4$ are invariant under $SL(2,{\bf R})$.
The associated orthonormal frame  is given by the following set of  1-forms,
\bea
\label{fullframe}
e ^i & = & f_4 \, \e ^i  \hskip 2in i=0,1,2,3
\no \\
e^4 & = & f_4 \, d\mu 
\no \\
e^5 & = & f_1 \, \e^5 = f_1 (d \beta +  A_1)
\no \\
e^a & = & f_2 \, \e ^a \hskip 2in a=6,7,8,9
\eea
For $i=0,1,2,3$, the $\e^i$ span the orthonormal frame for $AdS_4$
and may be chosen as follows,
\bea
\label{AdS4frame}
\e^0 = r^{-1} dr   \hskip 1in  \e^i = r^{-1}  dx^i  \qquad i=1,2,3
\eea
For $a=6,7,8,9$, the $\e^a$ span the orthonormal frame on $CP_2$ of (\ref{CP2frame}). The  volume form\footnote{We
shall introduce the following notation, $e^{i_1 i_2 \cdots i_p}
\equiv e^{i_1} \wedge e^{i_2} \wedge \cdots \wedge e^{i_p}$
and use it throughout.} is 
\bea
e^{0123456789} =
f_1 f_2 ^4 f_4^5 \, d \mu \wedge \e ^{0123} \wedge \e^{56789}
\eea

\subsection{The Ansatz for the antisymmetric tensor fields}

Invariance under  $SO(2,3) \times SU(3) \times U(1)_\beta $ requires the 
self-dual 5-form to be of the form,
\be
\label{ansatzF5}
F_{(5)}= f_5\; \left ( -  e^{01234} +    e^{56789}  \right ) \label{fivean}
\ee
where $f_5$ is a scalar function that depends only on $\mu$, by the same
arguments as used for the metric. Recall that $F_{(5)}$ and thus $f_5$
must  also be invariant under $SL(2,{\bf R})$.

\smallskip

To construct a 2-form $B_{(2)}$ which is invariant under 
$SO(2,3) \times SU(3) \times U(1)_\beta \times SL(2,{\bf R})$, we make use 
of the $SU(3)$ invariant 2-forms $K$, $\hat A_2$, and $\bar {\hat A_2}$.
Since $B_{(2)}$ is complex, we include $\hat A_2$ and $\bar {\hat A_2}$ 
with independent complex coefficient functions $f_3$ and $\bar g_3$, 
\bea
\label{fdefa}
B_{(2)} = i f_3 \hat A_2 -i  \bar g_3 \bar { \hat A_2} + f_6 K
\eea
It will turn out that the Type IIB field equation (\ref{Geq}) for $G_{\mu \nu \rho}$
force $f_6'=0$ and render this term pure gauge; therefore we shall set $f_6=0$ 
in the sequel.  The field strength $F_{(3)}$ is then, 
\bea
F_{(3)} =
i {f_3 ' \over f_4 f_2^2} e^4 \wedge A_2
-i  {\bar g _3 ' \over f_4 f_2^2} e^4 \wedge \bar A_2
- 3 {f_3 \over f_1 f_2^2} e^5 \wedge A_2
- 3 {\bar g _3 \over f_1 f_2^2} e^5 \wedge \bar A_2 
\eea
The associated composite $G$ is given by
\bea
\label{ansatzG}
G =
a \, e^5 \wedge A_2 - i b \, e^4 \wedge A_2 
+  c \, e^5 \wedge \bar A_2 -  i d \, e^4 \wedge \bar A_2
\eea
where the coefficient functions are given by
\bea
\label{abcd}
a = -{3\over  f_1 f_2^2}f( f_3-Bg_3)
& \hskip .7in &
c = -{3\over  f_1 f_2^2}f(\bar g_3-B \bar f_3)
\no \\
b = - { 1 \over  f_4 f_2^2}f(f_3'-Bg_3')
& \hskip .7in &
d = + {1 \over  f_4f_2^2}f(\bar g_3'-B \bar f_3')
\eea
For later convenience, we have here expressed these forms in terms of the 
frame $e^a$, for which
\bea
A_2 = f_2 ^2 \hat A_2 =  (e^6 + i e^7 ) \wedge (e^8 + i e^9) \, e^{3 i \beta}
\eea
With $f_3$ and $g_3$ functions only of $\mu$, the Ansatz in (\ref{fdefa}) is
invariant under $SO(2,3) \times SU(3)$. Under $U(1)_\beta$, 
the forms $A_2$ and $\bar A_2$ transform with constant opposite phases. 
Thus, $U(1)_\beta$ is not a symmetry of any one Ansatz, but rather relates
one Ansatz to another.

\subsection{Transformation properties under $SL(2,{\bf R})$}

Under $SL(2,{\bf R})$, the metric (in the Einstein frame) and the 5-form
$F_{(5)}$ are invariant. The dilaton/axion field $B$, and the associated function $f$, transform as
\bea
B^s = {u B + v \over \bar v B + \bar u} 
\hskip .7in 
f^s = |\bar v B + \bar u| \, f
\hskip .7in 
\left ( f^2 B' \right )^s = e^{2 i \theta } f^2 B'
\eea
where the superscripts $s$ indicate the transformed objects, $u,v \in {\bf C}$
and $\bar u u - \bar v v =1$.
The functions $f_3$ and $g_3$ transform linearly, according to (\ref{stransf}), 
and we have,
\bea
f_3 ^s = u f_3 + v g_3 
& \hskip 1in & 
f^s (f^s _3 - B ^s g_3 ^s ) = e^{i \theta} f (f _3 - B  g_3  )
\no \\
g_3 ^s = \bar v f_3 + \bar u g_3
& \hskip 1in &
f^s (\bar g ^s _3 - B ^s \bar f_3 ^s ) = e^{i \theta} f (\bar g  _3 - B  \bar f _3  )
\eea
where the phase $\theta$ is a field-dependent transformation parameter, given by
\bea
e^{ i \theta} = \left ( { v \bar B + u \over \bar v B + \bar u} \right )^\half
\eea
Since the phases of all terms in $G$ are the same, we get a covariant
formula, $G^s = e^{i \theta} \, G$. 

\newpage

%%%%%%%%%%%%%%%%%%%%%%%%%%%%%%%%%%%%%%%%%%
%%%%%%%%%%%%%%%%%%%%%%%%%%%%%%%%%%%%%%%%%%
\section{The Reduced Bianchi identities and Field Equations}
\setcounter{equation}{0}
%%%%%%%%%%%%%%%%%%%%%%%%%%%%%%%%%%%%%%%%%%
%%%%%%%%%%%%%%%%%%%%%%%%%%%%%%%%%%%%%%%%%%

In this section, we reduce the Type IIB Bianchi identities and field 
equations, given in section~2, to the Ansatz constructed in section 5.
The Bianchi identities (\ref{bianchi1}), (\ref{bianchi2}), and (\ref{bianchi3})
are automatically satisfied.  The Bianchi identity (\ref{bianchi4}) for $F_{(5)}$
reduces to
\bea
\label{f5eq}
f_5'= - 4 {f_2'\over f_2} f_5 -{f_1'\over f_1} f_5  
+ \half  f_4( a  \bar b +  \bar a  b + c \bar d + \bar c  d)
\eea
and is solved by
\bea
\label{bi1}
f_5 = { 3 \over 2} \, { |f_3|^2 - |g_3|^2 + C_1 \over f_1 f_2^4}
\eea
which fixes $f_5$ in terms of $f_1,f_2,f_3, g_3$, and the (real) integration
constant $C_1$.

\subsection{The reduced field equations for $B$ and $G$}

The field equation for the complex scalar $B$ reduces to
\bea
\label{eqB}
B'' + B' \bigg ( 3 {f_4'\over f_4} + {f_1'\over f_1}+ 4 {f_2'\over f_2} \bigg )
+ 2 f^2 \bar B B'B' + 2 {f_4 ^2 \over f^2} (ac-bd) =0
\eea
where the variables $a,b,c,d$ were introduced in (\ref{abcd}).

\smallskip

To reduce the field equations of the antisymmetric tensor field $B_{(2)}$,
it is convenient to first recast (\ref{Geq}) in terms of differential 
forms,\footnote{Our conventions for the Poincar\'e dual are given via the 
following pairing relation between two arbitrary rank $p$ differential
forms $S_{(p)}$ and $T_{(p)}$, by  $S_{(p)} \wedge *T_{(p)} 
= {1 \over p!} S_{(p)}^{a_1 \cdots a_p} T_{{(p)}a_1 \cdots a_p} e^{0123456789}$. 
In particular, we have $** S_{(p)}= (-1)^{p+1} S_{(p)}$, and the duals
$* e^{01234} = - e^{56789}$, and $*e^{6789} = e^{012345}$, which will be useful later on.}
\bea
 * d(* G) + i \,   (i_Q G) + (i_P \bar G) - 4 i \, (i_G F_{(5)}) =0
\eea
Here $i_V G$ stands for the contraction of $G$ with $V$.
To calculate this equation, it is helpful to have the following contractions,
$i_{e^5 \wedge A_2} \, e^{56789}  = A_2$, as well as the complex conjugate 
relation. Identifying terms in $A_2$, we find, after some simplification, 
\bea
\label{f3g3}
&&
f_3 '' - B g_3 '' - 9 {f_4^2 \over f_1^2} (f_3 - Bg_3) 
+ \left ( { f_1' \over f_1} + 3{ f_4' \over f_4} \right ) (f_3 ' - B g_3')
\no \\ && \hskip 1in
- 2 f^2 B' (g_3 ' - \bar B f_3') - 12 { f_4^2 f_5 \over f_1} (f_3 - B g_3) =0
\eea
while in $\bar A_2$, the equation is obtained from (\ref{f3g3})
by letting $f_3 \to \bar g_3$, and $f_5 \to - f_5$, leaving all other
functions unchanged. It is actually more convenient to express
these field equations in terms of the coefficient functions $a,b,c,d$ of $G$,
and we find (with $Q = Q_\mu d\mu$), 
\bea
\label{eqabcd}
a' -  i Q_\mu  a 
&=& -\Big( {f_1'\over f_1}+2 {f_2'\over f_2}\Big) a + 3{f_4\over f_1} b  - f^2 B' \bar c
\no \\
b' -  iQ_\mu  b
&=& -\Big( 4{f_4' \over f_4}+2{f_2'\over f_2} +{f_1'\over f_1}\Big)  b 
+3 {f_4 \over f_1} a - f^2 B' \bar d + 4 f_4 f_5 a
\no \\
c' -  i Q_\mu   c
&=& - \Big( {f_1'\over f_1}+2 {f_2'\over f_2}\Big) c
- 3 {f_4\over f_1} d  - f^2 B' \bar a
\no \\
d' -  i Q_\mu   d
&=&  -\Big( 4{f_4' \over f_4}+2{f_2'\over f_2}+{f_1'\over
f_1}\Big) d- 3 {f_4\over f_1} c +  4 f_4 f_5  c  - f^2 B' \bar b
\eea

\subsection{Reducing Einstein's equations}

To reduce the Einstein equations in (\ref{Eeq}), we first obtain the 
Ricci  curvature tensor. It is convenient to carry out all calculations
using the orthonormal frame of (\ref{fullframe}), which we shall denote 
collectively by $e^A$ where $A= (i,4,5,a)$ with $i=0,1,2,3$, and $a=6,7,8,9$.
The torsion-free connection $\omega ^A {}_B$, the associated curvature
$\Omega ^A {}_B$, the Riemann tensor $R^A{}_{BCD}$, and the Ricci tensor 
(all expressed in frame indices) are then defined by the relations,
\bea
\label{geometry}
0 & = & d e^A + \omega ^A {}_B \wedge e^B 
\no \\
\Omega ^A {}_B  & = & d \omega ^A {}_B + \omega ^A {}_C \wedge \omega ^C {}_B
\no \\
\Omega ^A {}_B & = &  \half R^A {}_{BCD} \, e^C \wedge e^D
\no \\
R_{BD} & = & R^A{}_{BAD}
\eea
where $A,B,C,D = 0,1,2,\cdots,9$. The corresponding objects for the unwarped $AdS_4$ and $CP_2$ geometries will be denoted with hats; they are 
the frames $\e^i, \e^a$, 
the connections  $\hat \omega ^i {}_j, \hat \omega ^a {}_b$, 
the curvatures  $\hat \Omega ^i {}_j, \hat \Omega ^a{}_b$, 
the Riemann tensors $\hat R^i {}_{jkl}, \hat R^a {}_{bcd}$,
and the Ricci tensors $\hat R_{jl}, \hat R_{bd}$. They all obey the equations
(\ref{geometry}) with hatted objects and for the ranges of indices 
$i,j,k,l =0,1,2,3$ and $a,b,c,d= 6,7,8,9$.
(The unwarped geometry also has structure in the direction 5, where it
can be viewed as the $S^1$ fiber of $S^5$ over $CP_2$; we shall 
treat this direction as separate.) The $AdS_4$ curvature $\hat \Omega$ 
is calculated using the unwarped frame in (\ref{AdS4frame}), and is
given by\footnote{The explicit forms of the $AdS_4$ and $CP_2$ 
unwarped connections will not be needed here.}
\bea
\hat \Omega _{ij} & = & 
- \half (\eta _{ik} \eta _{jl} - \eta _{il} \eta _{jk} ) \, \e^k \wedge \e^l
\no \\
\hat R_{ijkl} & = & -(\eta _{ik} \eta _{jl} - \eta _{il} \eta _{jk} )
\no \\
\hat R _{jl} & = &  - 3 \eta _{jl}
\eea
The $CP_2$  curvature is calculated using the unwarped frame in 
(\ref{CP2frame}) and is given by
\bea
\hat \Omega _{ab}  & = &
\left ( \delta _{ac} \delta _{bd} + k_{ac} k_{bd} +  k_{ab} k_{cd} \right )
\, \e^c \wedge \e^d
\no \\
\hat R _{abcd} & = & 
\delta _{ac} \delta _{bd}  - \delta _{ad} \delta _{bc} 
+ k_{ac} k_{bd} - k_{ad} k_{bc}  + 2  k_{ab} k_{cd} 
\no \\
\hat R_{bd} & = &  6 \delta _{bd}
\eea
where the tensor $k$ is anti-symmetric and defined by 
$k_{cd} \e^c \wedge \e^d = 2K = 2\e^6 \wedge \e^7 + 2 \e^8 \wedge \e^9$.

\smallskip

The connection components of the full geometry are needed to compute the 
curvature as well as the supersymmetry variation equations. 
They are given by $\omega _{AB} = - \omega _{BA}$ and
\bea
\label{connection}
\omega _{ij} =  \hat \omega _{ij} 
\hskip .3in 
& \hskip 1in &
\omega _{i5}  =0
\no \\
\omega ^i {}_4 =  {f_4' \over f_4^2} \, e^i
\hskip .15in
& \hskip 1in &
\omega _{ib} =  0
\no \\
\omega ^5 {}_4 =  {f_1 ' \over f_1 f_4 } \, e^5
& \hskip 1in &
\omega _{5a} =   {f_1 \over f_2^2} \, k_{ab} \, e^b
\no \\
\omega ^a {}_4 = {f_2' \over f_2 f_4} e^a
& \hskip 1in &
\omega _{ab} = \hat \omega _{ab} - {f_1 \over f_2^2} \, k_{ab} \, e^5
\eea
The components of the curvature form $\Omega ^A{}_B$ and of the 
Riemann tensor $R^A{}_{BCD}$ are needed only to evaluate the 
Ricci tensor, and will not be exhibited here. The non-vanishing 
components of the Ricci tensor are given by,
\bea
R_{ij} & = &
- \eta _{ij} \left ( {3 \over f_4^2} + 2 {(f_4')^2 \over f_4 ^4} + {f_4 '' \over f_4^3} 
+ {f_1' f_4' \over f_1 f_4^3} + 4 {f_2' f_4' \over f_2 f_4^3} \right )
\no \\
R_{44} & = &
- 4 {f_4 '' \over f_4^3} + 4 {(f_4')^2 \over f_4 ^4} + {f_1' f_4' \over f_1 f_4^3}
- {f_1'' \over f_1 f_4^2} + 4 {f_2' f_4' \over f_2 f_4^3}
- 4 {f_2'' \over f_2 f_4^2}
\no \\
R_{55} & = &
- {f_1 '' \over f_1 f_4^2} - 3  {f_1' f_4' \over f_1 f_4^3}
- 4 {f_1' f_2 ' \over f_1 f_2 f_4^2}  + 4 {f_1^2 \over f_2^4}
\no \\
R_{ab} & = &
\delta _{ab} \, \left ( { 6 \over f_2^2} - {f_2'' \over f_2 f_4^2} 
- 3 {f_2 ' f_4 ' \over f_2 f_4^3} - {f_1 ' f_2' \over f_1 f_2 f_4^2}
- 2 {f_1^2 \over f_2^4} - 3 {(f_2')^2 \over f_2 ^2 f_4^2} \right )
\eea
Finally, for later use, we record also the Ricci scalar,
\bea
R & = & {4 \over f_4^2} \left \{
3 {(f_2')^2 \over f_2 ^2 } + 3  {(f_4')^2 \over f_4 ^2} - 3
+ 2  {f_1 ' f_2' \over f_1 f_2 }
+ 2  {f_1' f_4' \over f_1 f_4} + 8 {f_2' f_4' \over f_2 f_4}
 -   {f_1^2 f_4^2 \over f_2^4} + 6 { f_4^2 \over f_2^2}  \right \}
\eea
which will be used to evaluate the reduced action.

\smallskip

It is straightforward to evaluate the ``matter" contributions to
Einstein's equations, which leads to the following reduced Einstein
equations.

\noindent
For $i,j=0,1,2,3$, we have,
\bea
\label{E1g}
&&
3  + 2 {(f_4')^2 \over f_4 ^2} + {f_4 '' \over f_4} 
+ {f_1' f_4' \over f_1 f_4} + 4 {f_2' f_4' \over f_2 f_4}
- 4  f_5^2 f_4^2 
\no \\ && \hskip 2in
+ \half f_4^2 \left (- |a|^2 - |b|^2 - |c|^2 - |d|^2 \right ) =0
\eea
For $44$, we have,
\bea
\label{E2g}
&&
4 {f_4 '' \over f_4} - 4 {(f_4')^2 \over f_4 ^2} - {f_1' f_4' \over f_1 f_4}
+ {f_1'' \over f_1 } - 4 {f_2' f_4' \over f_2 f_4}
+ 4 {f_2'' \over f_2 } - 4 f_5^2 f_4^2 +  2 f^4 |B'|^2  
\no \\ && \hskip 2in
+ \half f_4^2 \left (- |a|^2 +3 |b|^2 - |c|^2 +3 |d|^2 \right ) =0
\eea
For $55$, we have
\bea
\label{E3g}
&&
{f_1 '' \over f_1 } + 3  {f_1' f_4' \over f_1 f_4}
+ 4 {f_1' f_2 ' \over f_1 f_2 }  - 4 {f_1^2 f_4^2 \over f_2^4}
+ 4 f_5^2 f_4^2
\no \\ && \hskip 2in
+ \half f_4^2 \left ( 3 |a|^2 - |b|^2 +3 |c|^2 - |d|^2 \right ) =0
\eea
and finally for $a,b=6,7,8,9$, we have,
\bea
\label{E4g}
&&
 -{ 6 f_4^2 \over f_2^2} + {f_2'' \over f_2 } 
+ 3 {f_2 ' f_4 ' \over f_2 f_4} + {f_1 ' f_2' \over f_1 f_2 }
+ 2 {f_1^2 f_4^2 \over f_2^4} + 3 {(f_2')^2 \over f_2 ^2 } 
+ 4 f_5^2 f_4^2
\no \\ && \hskip 2in
+ \half f_4^2 \left ( |a|^2 + |b|^2 + |c|^2 + |d|^2 \right ) =0
\eea
Recall that here, as before, the variables $a,b,c,d$ are given in terms 
of the independent functions $f_3,g_3$ by the definitions (\ref{abcd}).

\subsection{Integral of motion}

The 4 Einstein equations possess a single integral of motion,
which may be viewed as the (vanishing) Hamiltonian,
or Wheeler-De Witt equation. It is obtained by adding
the 4 equations with the following multiplicative factors, $+4, -1, +1, +4$,
and is given by
\bea
\label{integral}
&&
 12 +  12  {(f_2')^2 \over f_2 ^2 } + 12  {(f_4')^2 \over f_4 ^2} 
+ 8  {f_1 ' f_2' \over f_1 f_2 }  + 8  {f_1' f_4' \over f_1 f_4}  
+ 32 {f_2' f_4' \over f_2 f_4} + 4  {f_1^2 f_4^2 \over f_2^4}
\no \\ && \hskip .4in 
 -24 { f_4^2 \over f_2^2} +8 f_4^2 f_5^2 - 2 f^4 |B'|^2   
 + 2 f_4^2 \left ( |a|^2 - |b|^2 + |c|^2 - |d|^2 \right ) =0
\eea
In general, the above system of reduced equations (including the reduced
equations for $B$ and for $a,b,c,d$) does not appear to
possess any further first integrals.

\subsection{Reduced Action Principle}

In this subsection, we shall show that the reduced field equations, listed above, 
may be derived from the Type IIB action (\ref{action1}), reduced to our Ansatz. 
To show this, we shall need one more ingredient in the construction of
the Ansatz that was not needed for the field equations, but is needed for the
action, namely the anti-symmetric tensor $C_{(4)}$. The starting point
is the relation between $C_{(4)}$ and $F_{(5)}$ in (\ref{GF5}). 
Using the Ansatz for $F_{(5)}$, $B_{(2)}$ and $F_{(3)}$, as well as
the Bianchi identities (\ref{bi1}), we obtain the following on-shell
expression for $dC_{(4)}$,
\bea 
\label{dC4}
d C_{(4)} =  - f_4 ^5 f_5 \, \hat e^{01234} + { 3 \over 2 } C_1 \hat e^{56789}  
- { i \over 4} \left ( f_3 \bar f_3 ' - \bar f_3 f_3 ' 
-g_3 \bar g_3' + \bar g_3 g_3 ' \right )  \, \hat e^{46789}
\eea
Therefore, the most general Ansatz for $C_{(4)}$ is as follows,
\bea
\label{C4}
C_{(4)} = g_5 \, \hat e^{0123} + { 3 \over 2} C_1 \, \beta \, \hat e^{6789}
+ h_5 \hat e^{6789}
\eea
where $g_5$ and $h_5$ are functions of $\mu$ only. The term proportional to
$C_1$ accounts for the term proportional to $C_1$ in (\ref{dC4}), via the
fact that $\hat e^{56789} = d\beta \wedge \hat e^{6789}$.

\smallskip

To obtain an off-shell formulation, for use in the action,  we postulate 
(\ref{C4})  as the Ansatz for $C_{(4)}$. This is clearly the most general 
$SO(2,3)\times SU(3) \times U(1)_\beta$ invariant 4-form we can write down.\footnote{Notice that under the action of $U(1)_\beta$, the form 
$C_{(4)}$ is not strictly invariant, but changes by a gauge transformation 
since  $\hat e^{6789}$ is a closed form.}
Evaluating $F_{(5)}$ now from the invariant Ansatz for $C_{(4)}$ in (\ref{C4}), 
we find an expression which is not necessarily self-dual,
\bea
F_{(5)} & = & g_5' \, \hat e^{01234} + { 3 \over 2} C_1  \hat e^{56789}
+ h_5 ' \hat e^{46789}
\no \\ &&
+ { i \over 4} \left ( f_3 \bar f_3 ' - \bar f_3 f_3 ' 
-g_3 \bar g_3' + \bar g_3 g_3 ' \right ) \hat e^{46789}
\no \\ &&
+ { 3 \over 2} \left ( |f_3|^2 - |g_3|^2 \right ) \hat e^{56789}
\eea
In turn, self-duality of $F_{(5)}$ requires the following relations,
\bea
\label{selfdual}
0 & = & f_4 ^{-5} g_5 ' + { 3 \over 2} { |f_3|^2 - |g_3|^2 + C_1 \over f_1 f_2^4}
\no \\
0 & = & h_5 ' + { i \over 4} \left ( f_3 \bar f_3 ' - \bar f_3 f_3 ' 
-g_3 \bar g_3' + \bar g_3 g_3 ' \right )
\eea
which reproduce the on-shell relations (\ref{dC4}).

\smallskip

We are now ready to reduce the Type IIB supergravity action of (\ref{action1}).
It will be expressed here in terms of our complex fields $B$, and $G$, and is given by, \footnote{Compared to the 
conventions of \cite{pol}, we omit an overall factor of $1/2\kappa _{10}^2$,
divide $C_{(4)}$ by a factor of 4, and divide $F_{(5)}$ by a factor of 4.
The relative normalization of the Chern-Simons term against the $|F_{(5)}|^2$
term is checked using the self-duality of $F_{(5)}$ and the Bianchi
identity for $F_{(5)}$. The absolute normalization of the $|F_{(5)}|^2$
term may be checked from Einstein's equations, while that of the 
Chern-Simons term may be checked independently against the 
field equation for $G$.}
\bea
S & = & 
\int dx \sqrt{g} \left \{
R - 2 f^4 \p_M B \p^M \bar B
- { 1 \over 12} G _{MNP} \bar G ^{MNP}
- 4 | F_{(5)}|^2 \right \}
\no \\ && \hskip .5in 
- i \int C_{(4)} \wedge G \wedge \bar G
\eea
Since, after variation of the fields, we are to further enforce the self-duality 
relation, this action is not quite unique. Indeed, we are free to add to the 
Lagrangian  density a term proportional to the square of the self-duality 
relation  $|F_{(5)} - * F_{(5)}|^2$, as its variation will vanish on self-dual fields. 
In reducing the  action over our Ansatz, it will be 
convenient to add the term $4|F_{(5)} - * F_{(5)}|^2$ to the Lagrangian density,
as this will eliminate terms in the reduction that are quartic in $f_3$ and $g_3$.

\smallskip

Evaluating this modified action on our Ansatz, omitting the overall volume factors 
of $\hat e^{0123} \wedge \e^{56789}$ of $AdS_4 \times S^5$, and integrating 
all second derivatives by part to convert all terms to involve only first 
derivatives, we obtain a 
\bea
S_{\rm reduced} & = &  \int d\mu \Bigg \{
12 (f_2')^2 f_1 f_2^2 f_4^3 
+ 12  (f_4')^2  f_1 f_2^4 f_4
+ 8  f_1 ' f_2'  f_2^3 f_4^3
+ 8  f_1' f_4'   f_2^4 f_4^2
\no \\ && \hskip .5in
+ 32 f_2' f_4' f_1 f_2^3 f_4^2
- 12 f_1 f_2^4 f_4^3 
- 4  f_1^3 f_4^5 
+ 24  f_1 f_2^2 f_4^5
 - 2 f^4 f_1 f_2^4 f_4^3 |B'|^2  
\no \\ && \hskip .5in
 - 2  f_1  f_4^3 \bigg (   f^2 |f_3' - B g_3' |^2 +  f^2 |\bar g_3' - B \bar f_3 ' |^2 \bigg )
\no \\ &&  \hskip .5in
- 18  { f_4^5 \over f_1  } 
	\bigg (  f^2 |f_3 - B g_3 |^2 +  f^2 |\bar g_3 - B \bar f_3|^2 \bigg )  
\no \\ && \hskip .5in
+ 8 f_1 f_2^4 f_4^{-5} (g_5')^2
+ 24 g_5 ' \left ( |f_3|^2 - |g_3|^2 \right )\Bigg \}
\eea
We have shown that the reduced field equations follow from the action 
$S_{\rm red}$. 

\smallskip

The only further relations implied by the self-duality constraint (\ref{SDeq})
are the value of the constant $C_1$ and the expression for the
function $h_5$ which yields  (\ref{selfdual}).

\newpage

%%%%%%%%%%%%%%%%%%%%%%%%%%%%%%%%%%%%%%%%%%
%%%%%%%%%%%%%%%%%%%%%%%%%%%%%%%%%%%%%%%%%%
\section{Supersymmetry variations}
\setcounter{equation}{0}
%%%%%%%%%%%%%%%%%%%%%%%%%%%%%%%%%%%%%%%%%%
%%%%%%%%%%%%%%%%%%%%%%%%%%%%%%%%%%%%%%%%%%

In this section, we reduce the BPS equations $\delta \lambda = \delta \psi _\mu =0$
expressing the vanishing supersymmetry variation of the dilatino and gravitino
fields. Fields satisfying these reduced equations with non-vanishing supersymmetry
parameter $\ep$ will exhibit some degree of residual supersymmetry. It will be convenient to express the gravitino equation in differential
form notation. The dilatino and gravitino equations are given respectively by,
\bea
\label{dilatino}
(\G \cdot P) {\cal B}^{-1} \epsilon^* 
- 
{1 \over 24} ( \Gamma \cdot G ) \, \epsilon & = & 0
\\
\label{gravitino1}
d \ep + \omega \, \ep + \varphi ^{(1)} \ep +  \varphi^{(2)} \B^{-1} \ep ^*  & = & 0
\eea
where the connection components are as follows,\footnote{Throughout, we shall
use the notation $\G \cdot T \equiv \G^{M_1 \cdots M_n} T_{M_1 \cdots M_n}$
for the contraction of rank $n$ totally anti-symmetric $\G$-matrices and tensors.}
\bea
\label{gravitino2}
\omega & = & {1 \over 4} \G^{MN}  \omega _{MN}
\no \\
\varphi ^{(1)} & = & 
 - {i \over 2} Q + {i \over 480} (\G \cdot F_{(5)} ) e_A  \G ^A 
\no \\
\varphi ^{(2)} & = &
- {1 \over 96} e_A \left ( \G^A (\G \cdot G) + 2 (\G \cdot G) \G^A \right )
\eea
Here, $e^A$ is the frame of (\ref{fullframe}), $\omega$ its torsion-free 
connection,  $\B$ is the complex conjugation matrix,  and we have used 
the following relation
\bea
\G ^{MNPQ} G_{NPQ} - 9 \eta ^{MN} \G^{PQ} G_{NPQ}
=
- \G^M (\G\cdot G) -2 (\G \cdot G) \G^M
\eea
to relate the gravitino equation of (\ref{susy1}) to that of (\ref{gravitino1})
and (\ref{gravitino2}).

\subsection{$\G$-matrices}

It will be useful to choose a well-adapted basis for 
$\G$-matrices,\footnote{A useful set of conventions for the $4 \times 4$ 
matrices $\gamma ^A$,
as well as some further formulas for combinations of $\G$-matrices, 
such as $\G ^{z_i}$, and $\G^{AB}$, are presented in Appendix B.}
\bea
\G ^\mu & = & \sigma _1 \otimes \gamma ^\mu \otimes I_4 \hskip 1in \mu = 0,1,2,3,4
\no \\
\G ^ m & = & \sigma _2 \otimes I_4 \otimes \gamma ^m \hskip 1in m=5,6,7,8,9
\eea
for which we have 
\bea
\label{chiralB}
\G^{11} & = &  \G^{0123456789} = \sigma _3 \otimes I_4 \otimes I_2 \otimes I_2  
\no \\
\B & = & \G^{2568} \hskip .35in = \sigma _3 \otimes \gamma ^2 \otimes \sigma _1 \otimes \sigma _2
\eea

\subsection{Reducing the dilatino equation}

Evaluating the dilatino equation (\ref{dilatino}) on the Ansatz, we find,
\bea 
4 { f^2 B' \over f_4} \, \G ^4  {\cal B}^{-1} \epsilon^* 
= 
\left ( a \G^5 -ib \, \G^4 \right ) \G^{z_1 z_2} e^{+3i\beta} \ep
+ \left ( c \G^5 - i d \, \G^4 \right )  \G ^{ \bar z_1 \bar z_2} e^{-3i\beta} \ep
\eea
To solve this equation, we multiply both sides to the left by $\G^4$
and use the fact that $\G^{45}$ commutes with $\G^{z_1 z_2}$.
The rhs then automatically projects onto the subspace of spinors
corresponding to eigenvalue $+1$ of the Dirac matrix $I_2 \otimes I_4 \otimes \gamma ^5$. Since $\B$ commutes with this matrix, both $\ep$ and $\B^{-1} \ep ^*$
must have eigenvalue $+1$. Introducing a basis of 2-component spinors 
$u_\pm$, with $\sigma _1 \, u_\pm = + u_\mp$, and  $\sigma _\pm \, u_\pm =0$,
we  parametrize of the eigenvalue $+1$ subspace for chiral spinors $\ep$ 
satisfying $\G^{11} \ep = - \ep$, in terms of two 4-dimensional complex spinors 
$\zeta _\pm$, and their  complex conjugates $\zeta _\pm ^*$,
\bea
\ep & = & u_- \otimes \Big ( 
\zeta _+ \otimes u_+ \otimes u_+ + \zeta _- \otimes u_- \otimes u_- \Big )
\no \\
\B ^{-1} \ep ^* & = & u_- \otimes \left ( 
i \gamma ^2 \zeta _-^*  \otimes u_+ \otimes u_+ 
- i \gamma ^2 \zeta _+^*  \otimes u_- \otimes u_- \right ) \label{cptwoans}
\eea
In this basis, the dilatino equation reduces to,
\bea
\label{dilatino2}
(a \gamma ^4 + b) \zeta _- & = & 
e^{-3i\beta} \, f^2 B'   (f_4)^{-1} \gamma ^2 \zeta _- ^*
\no \\
(c \gamma ^4 + d) \zeta _+ & = &
e^{+ 3 i \beta}  \, f^2 B' ( f_4)^{-1} \gamma ^2 \zeta _+ ^*
\eea
Notice that the equations for $\zeta _+$ and $\zeta _-$ can be satisfied 
independently of one another. Non-trivial solutions require the following relations,
\bea
\label{dilatino3}
\zeta _+ \not= 0 & \hskip .3in & 
(c + d ) (\bar c - \bar d) =  f^4 |B'|^2 ( f_4)^{-2}
\no \\
\zeta _- \not= 0 & \hskip .3in & 
(a+b ) (\bar a - \bar b) = f^4 |B'|^2 ( f_4)^{-2}
\eea
The $\beta$-dependence of $\zeta _\pm$ is readily solved for,
since $a,b,c,d$, $f_4$ and $f^2B'$ are all $\beta$-independent. 
The general solution is given by
\bea
\label{zetahat}
\zeta _\pm = e^{\pm {3\over 2} i \beta} \hat \zeta _\pm
& \hskip 1in & 
(a \gamma ^4 + b) \hat \zeta _- =  f^2 B' ( f_4)^{-1} \gamma ^2 \hat \zeta _- ^* 
\no \\ &&
(c \gamma ^4 + d) \hat \zeta _+ = f^2 B' (f_4)^{-1} \gamma ^2 \hat \zeta _+ ^* 
\eea
where $\hat \zeta _\pm$ is independent of $\beta$.

\subsection{Reducing the gravitino equation}

We begin by decomposing the gravitino equations (\ref{gravitino1})
onto the eigenvalue $+1$ and $-1$ of the matrix $\G^{6789} = 
- I_8 \otimes \sigma _3 \otimes \sigma _3$. Since $\ep$ and $\B^{-1} \ep ^*$
belong to the subspace with definite eigenvalue $-1$, this decomposition 
may be achieved by decomposing the covariant derivative according to 
components that commute or anti-commute with $\G^{6789}$.
This decomposition is unique since $\G^{6789}$ is invertible. We have,
\bea
\label{gravitino3}
d \ep + \omega_+ \, \ep + \varphi ^{(1)} _+ \, \ep +  \varphi ^{(2)} _+ \B^{-1} \ep ^* 
& = & 0
\no \\
\omega _- \, \ep + \varphi ^{(1)} _- \, \ep +  \varphi ^{(2)} _- \B^{-1} \ep ^* 
& = & 0
\eea
Here, the components that commute with $\G^{6789}$ are given by,
\bea
\omega_+  & = & 
{1 \over 4} \omega _{ij} \G^{ij} 
+ \half \omega _{i4} \G^{i4} 
+ \half \omega _{45} \G^{45}
+ {1 \over 4} \omega _{ab} \G^{ab}
\no \\
\varphi ^{(1)} _+   & = & 
- {i \over 2} Q 
+ {i \over 480} (\G \cdot F_{(5)} ) \left ( e_i \G ^i  + e_4 \G^4 + e_5 \G^5 \right )
\no \\
\varphi ^{(2)} _+ & = &
- {1 \over 96} \sum _{A=0,1,2,3,4,5}
e_A \left ( \G^A (\G \cdot G) + 2 (\G \cdot G) \G^A \right )
\eea
while the components that anti-commute with $\G^{6789}$ are given by
\bea
\omega _- & = &  \half  \omega _{a4} \G^{a4} + \half  \omega _{a5} \G^{a5}
\no \\
\varphi ^{(1)} _-   & = & 
 {i \over 480} (\G \cdot F_{(5)} ) e_a \G ^a
\no \\
\varphi ^{(2)}_- & = &
- {1 \over 96} e_a \left ( \G^a (\G \cdot G) + 2 (\G \cdot G) \G^a \right )
\eea
Here, the indices $i,j$ continue to run over the range $0,1,2,3$, while
$a,b$ run over $6,7,8,9$.

\subsubsection{Reducing the algebraic equation}

The second equation in (\ref{gravitino3}) is purely algebraic in $\ep$.
Using the expressions for the connection components of (\ref{connection}),
and the anti-symmetric tensor fields $F_{(5)}$ of (\ref{ansatzF5})
and $G$ of (\ref{ansatzG}), we find one equation involving both $\zeta _+$
and $\gamma ^2 \zeta _+^*$, and another equation involving both $\zeta _-$
and $\gamma ^2 \zeta _- ^*$. 
% ED revision begin
We proceed by assuming that $B'\not= 0$
since we shall be interested in obtaining Janus solutions with varying dilaton field. 
We may then eliminate  
% ED revision end
$\gamma ^2 \zeta _\pm^*$ using the dilatino equation 
(\ref{dilatino2}). The final result is expressed in terms of $\hat \zeta _\pm$,
using (\ref{zetahat}), by the following two independent equations,
\bea
\left ( 2 { f_2' \over f_2 } \gamma ^4 
\pm  2 {f_1 f_4 \over f_2^2}  - 2 f_5 f_{4}
- { f_4^2 \over f^2 B'} \gamma ^4 
(a +  b \gamma ^4) (c +  d \gamma ^4) \right ) 
\hat \zeta_\pm = 0
\eea
where $\pm$ is correlated with the subscript of $\hat \zeta _\pm$.

\smallskip

The above equations imply that, still under the assumption that $B'\not= 0$,
either $\hat \zeta _+$ or $\hat \zeta _-$ (or both) must vanish. To show this, 
we begin by assuming that $\hat \zeta _+\not=0$, and decompose 
the equation for $\hat \zeta _+$ onto the two $\gamma ^4$ chiralities
of $\hat \zeta _+$. Note that the equations must hold for both chiralities,
since the dilatino equation (\ref{dilatino2}) chirality,
\bea
\label{algeq}
+ 2 { f_2' \over f_2 }  +  2 {f_1 f_4 \over f_2^2} 
- 2 f_5 - {  f_4 ^2 \over f^2 B'}  (a + b) (c + d) & = & 0
\no \\
- 2 { f_2' \over f_2 }  +  2 {f_1 f_4 \over f_2^2} 
- 2 f_5 + {  f_4 ^2 \over f^2 B'}  (a - b) (c - d) & = & 0
\eea
where the $\pm $ sign refers to the $\gamma ^4$-chirality of $\hat \zeta _+$.
Using these two relations, required by $\hat \zeta _+ \not=0$ in the 
equation for $\hat \zeta _-$ gives $f_1 f_4  f_2^{-2} \hat \zeta _-=0$.
For non-degenerate geometries,  $f_1f_4  f_2^{-2} \not=0$ and thus
we must have $\hat \zeta _-=0$, as announced. 

\smallskip

The equations for $\hat \zeta _+$ and $\hat \zeta _-$ are mapped into
one another by parity, which maps $\mu \to - \mu$, and are
equivalent to one another. In view of the result above, we may choose
$\hat \zeta _- =0$.

\subsubsection{Reducing the differential equation}

To reduce the differential equation in (\ref{gravitino3}), we begin by 
calculating the connection $\omega _{ab} \G^{ab}$, using $\G^{ab}
= I_8 \otimes \gamma ^{ab}$. Furthermore, we use the  relations 
\bea
\gamma ^{67} u_\pm \otimes u_\pm = 
+ \gamma ^{89} u_\pm \otimes u_\pm & = & \pm i u_\pm \otimes u_\pm
\no \\
\gamma ^{68} u_\pm \otimes u_\pm = 
- \gamma ^{79} u_\pm \otimes u_\pm & = & \pm  u_\mp \otimes u_\mp
\no \\
\gamma ^{69} u_\pm \otimes u_\pm = 
+ \gamma ^{78} u_\pm \otimes u_\pm & = &  i \, u_\mp \otimes u_\mp
\eea
and the fact that $\hat \omega _{68} = \hat \omega _{79}$, and 
$\hat \omega _{69} = - \hat \omega _{78}$, as found in
(\ref{CP2connection}), to derive
\bea
\label{newA1}
{1 \over 4} \omega _{ab} \gamma ^{ab}
= { 3 i \over 2} A_1 \sigma _3 -i {f_1 \over f_2^2} e^5 \sigma _3 
\eea
Here, $\sigma _3$ acts on $\zeta _\pm$ by 
$\sigma _3 \zeta _\pm = \pm  \zeta _\pm$. 
All  connection components now commute with $\sigma _3$ and
we may decompose the reduced gravitino equation onto its decoupled  
$\zeta _+$  and $\zeta _-$ components. Finally, the simple 
$\beta $-dependence of $\zeta_\pm$ found in (\ref{zetahat}) is used
to recast the equations in terms of $\hat \zeta _\pm$. The differential
$d\beta$ picked up in the process combines with the $U(1)$ connection 
$A_1$ in (\ref{newA1}) and forms the gauge invariant combination 
$d\beta + A_1 = \hat e^5$, so that we have effectively
\bea
{1 \over 4} \omega _{ab} \gamma ^{ab}
 \, \to \, 
  { 3 i \over 2f_1} e^5 \sigma _3 -i {f_1 \over f_2^2} e^5 \sigma _3 
\eea
Finally, retaining only the $\hat \zeta _+$-component of the differential
equation, we obtain
\bea
\label{reddiff}
0 & = &
d \hat \zeta _+ + \left ( 
{1 \over 4} \hat \omega _{ij} \gamma ^{ij}   + \half \omega _{i4} \gamma ^{i4} 
- {i \over 2}  \omega _{45} \gamma ^4
+{ 3 i \over 2f_1 } e^5  -i {f_1 \over f_2^2} e^5  \right ) \hat \zeta _+
\no \\ && \hskip .3in 
+ \left ( - {i \over 2} Q 
+ \half f_5  (  e_i \gamma  ^i  + e_4 \gamma ^4 - i e_5   ) \right ) \hat \zeta_+
\no \\ && \hskip .3in
+ {1 \over 4} \bigg  (
 e_i \gamma ^i  (a + b \gamma ^4) 
+ e_4 \gamma ^4  (a - 3 b \gamma ^4) 
+ e_5  (-3i a +i b \gamma ^4)
\bigg ) \gamma ^2 \hat \zeta ^*_+ \qquad
\eea

\subsubsection{Analyzing the reduced differential equation}

Since $\hat \zeta _+$ is  $\beta$-independent in (\ref{reddiff}),
no $ \p_\beta \hat \zeta _+ d \beta$ contribution will appear in $d\hat \zeta _+$.
As a result, we must have $i_{e^5} d \hat \zeta _+ =0$, which yields a further algebraic equation, 
\bea
\label{e5eq}
\left (
{f_1' \over f_1 } \gamma ^4 + 3 { f_4 \over f_1} - 2 {f_1 f_4 \over f_2^2} 
- f_4 f_5 \right ) \hat \zeta _+
=
\half f_4 (3 a -  b \gamma ^4 ) \gamma ^2 \hat \zeta ^* _+
\eea
The decomposition of (\ref{reddiff}) onto the directions $e^4$ and $\e_i$,
yields two differential equations,
\bea
\big( \hat \nabla ^i +{1\over2}{f_4'\over f_4}\gamma^i \gamma_4 
+{1\over 2} f_5f_4\gamma^i \big) \hat \zeta_+ 
+  {1 \over 4} f_4( a-  b \gamma_4)\gamma^i\gamma_2 \hat \zeta_+^*
&=& 0
\\
\big(\partial_\mu-{i\over 2}Q + {1\over 2} f_5 f_4 \gamma_4\big) \hat \zeta_+ 
+ {1 \over 4} f_4( a \gamma_4- 3 b)\gamma_2 \hat \zeta_+^*
&=&0
\label{dmuder}
\eea
where $\hat \nabla ^i$ is the covariant derivative with connection 
$\hat \omega _{ij} \gamma ^{ij}/4$. Finally, we proceed to eliminating
the spinor $\gamma ^2 \hat \zeta _+^{*}$ using the dilatino equation
(\ref{dilatino2}), and to further simplifying the equations. 
From the outset, we use the fact that $\hat \zeta _-=0$, as derived earlier.
The results for all supersymmetry variation equations are summarized below.

\subsection{Summary of the reduced dilatino/gravitino equations}

\noindent
$\bullet$ The implications of the dilatino equations,
\bea
\label{eq11}
(c +  d) ( \bar c - \bar d) & = & f^4 |B'|^2 ( f_4)^{-2} 
\eea
$\bullet$ The algebraic integrability equations,
\bea
\label{eq21}
{ f_2' \over f_2}  & = & { f_4 ^2 \over 2 f^2 B'}   ( a c + b d)
\\
\label{eq22}
{f_1  f_4 \over f_2^2}  -  f_4 f_5 
& = &   { f_4 ^2 \over 2 f^2 B'} ( a d + b  c)  
\eea
$\bullet$ The $e^5$ component equation,
\bea
\label{eq31}
{f_1'\over f_1} 
& = &  {  f_4 ^2 \over 2 f^2 B'} (3 a c - b  d)
\\
\label{eq32}
 3 {f_4\over f_1} - 2 {f_1f_4\over f_2^2}- f_4 f_5
& = &   {  f_4 ^2 \over 2 f^2 B'} (3 a d - b c ) 
\eea
$\bullet$ The $AdS_4$ components' equation,
\bea
\label{eq4}
 {f_1^2   f_4^2  \over f_2^4}  
- \left ( { f_2' \over f_2}  + { f_4' \over f_4}  \right )^2 =1
\eea

\smallskip

This summary of results is obtained as follows. Equation (\ref{eq11})
is just the original dilatino equation of (\ref{dilatino3}) for the case
where $\hat \zeta _-=0$. Equations (\ref{eq21}) and (\ref{eq22}) are 
obtained by taking the sum and the difference of the two
algebraic equations (\ref{algeq}). Equations (\ref{eq31}) and (\ref{eq32})
are obtained (\ref{e5eq}) by eliminating $\gamma ^2 \hat \zeta _+^{*}$
using the dilatino equation (\ref{dilatino2}), and taking the sum and 
difference of the resulting equations for $\pm 1$ eigenvalues of 
the $\gamma ^4$ matrix. To obtain equation (\ref{eq4}) requires more 
work. We begin by eliminating the spinor $\gamma ^2 \hat \zeta _+^{*}$ using
the dilatino equation (\ref{dilatino2}). The resulting equation takes the form,
\bea
\label{ads4}
\left ( \hat \nabla _i + \kappa \gamma _i + \lambda \gamma _i \gamma ^4 \right )
\hat \zeta _+ =0
\eea
where
\bea
\kappa & = & 
\half f_4 f_5 +  {  f_4 ^2 \over 4 f^2 B'} (a d +  b  c)  
\hskip .3in =
\half \, {f_1 f_4 \over f_2^2}
\no \\
\lambda & = & 
\half {f_4' \over f_4} + {  f_4 ^2 \over 4 f^2 B'} ( a  c +  b d)
\hskip .4in = 
\half {f_2 ' \over f_2} + \half {f_4' \over f_4}
\eea
In deriving the second equalities on the rhs of the equations above,
we have eliminated the combination $ad+bc$ using (\ref{eq22}), and 
eliminated the combination $ac+bd$ using (\ref{eq21}). 
The coefficients $\kappa$ and $\lambda$ depend upon $\mu$
but not upon the variables of $AdS_4$, and may thus be viewed 
as constants in this differential equations.

\smallskip

The integrability of (\ref{ads4})
requires $\kappa ^2 - \lambda ^2 = 1/4$, which yields (\ref{eq4}).
Finally, equation (\ref{dmuder}) is not reproduced in the summary
because it only governs the $\mu$-evolution of $\hat \zeta _+$,
is always integrable, and imposes no new condition of the dynamical variables 
of the system, namely $f_1,f_2,f_3,g_3,f_4$ and $B$.

\smallskip

% revision MG %
The solution of equations (\ref{eq11}) to (\ref{eq4}) gives a background which preserves 
four real supercharges. The counting proceeds as follows: In Type IIB supergravity the supersymmetry transformation parameter $\epsilon$  has 32 real components. 
The compatibility with the $CP_{2}$ fibration leads to (\ref{cptwoans}) reducing 
$\epsilon$ to 16 real components parameterized by two complex four dimensional 
spinors $\zeta_{+}$ and $\zeta_{-}$. The analysis of section 7.3 implies that one of 
the two $\zeta_{\pm}$ is zero, leaving four complex components. Finally the dilatino supersymmetry condition leads to a reality condition leaving four real unbroken supersymmetries, which is the degree of supersymmetry of our solution. 
In the sequel, we shall not need to explicitly solve for the spinor $\hat \zeta_{+}$;  
instead it will suffice to solve the integrability conditions which guarantee the existence 
of the four real unbroken supersymmetry.
% end revision %

\newpage

%%%%%%%%%%%%%%%%%%%%%%%%%%%%%%%%%%%%%%%%%%
%%%%%%%%%%%%%%%%%%%%%%%%%%%%%%%%%%%%%%%%%%
\section{Reality properties of supersymmetric solutions}
\setcounter{equation}{0}
%%%%%%%%%%%%%%%%%%%%%%%%%%%%%%%%%%%%%%%%%%
%%%%%%%%%%%%%%%%%%%%%%%%%%%%%%%%%%%%%%%%%%

We shall now search for supersymmetric solutions. This requires
satisfying the field equations (\ref{eqB}), (\ref{eqabcd}), 
and  (\ref{E1g}),  (\ref{E2g}),  (\ref{E3g}),  (\ref{E4g}), (one special 
combination of which is the constraint (\ref{integral})), as well as the 
supersymmetry conditions (\ref{eq11}--\ref{eq4}) of the summary. 
Assuming that all these equations are simultaneously  
satisfied guarantees that we will have a true solution to the field 
equations, which also is supersymmetric. The simultaneous consideration
of the susy equations and the field equations leads to many simplifications.

\subsection{Solving the equations for $B$ and $\tau$}

The difference between (\ref{eq21}) and (\ref{eq31}) yields
\bea
{f_1 ' \over f_1} - {f_2' \over f_2} = { f_4^2 \over f^2 B'} (ac-bd)
\eea
which upon eliminating $ac-bd$ in (\ref{eqB}), and dividing the resulting 
equation by $B'$ gives
\bea
\label{eqB1}
0 = {B'' \over B'} +3  {f_1 ' \over f_1} +2  {f_2' \over f_2}  + 3{f_4 ' \over f_4} 
+ 2 f^2 \bar B B'
\eea
Taking the real part of (\ref{eqB1}), and using the fact that 
$f^2 (\bar B B' + B \bar B ') =  (\ln f^2)'$,  we get 
\bea
f^2 |B'| = {  C_2  \over f_1 ^3 f_2 ^2 f_4^3} \label{dileqa}
\eea
where $C_2$ is a constant. Converting the imaginary part of (\ref{eqB1}),
into an equation for $\tau= \tau _1 + i \tau _2$, using (\ref{Btau}), we get
precisely the imaginary part of the $\tau$-equation for the non-supersymmetric
Janus solution of (\ref{taueq}),
\bea
0= { \tau '' \over \tau'} - { \bar \tau '' \over \bar \tau'}
+ 2 i { \tau _1 ' \over \tau _2} 
\eea
Just as in the case of the non-supersymmetric Janus solution, its 
solution requires the axion/dilaton field $\tau$ to flow along a 
geodesic in the $\tau$-upper-half-plane,
\bea
|\tau - p|^2 = r^2 
\eea
where $p,r$ are arbitrary real constants. 

\subsection{Mapping to real solutions}

Since $B$ flows along a geodesic, and $SL(2,{\bf R})$ acts transitively
on the space of all  geodesic segments of the same hyperbolic 
length,\footnote{By a geodesic segment we understand a connected
segment of a single geodesic.}
we may use an $SL(2,{\bf R})$ transformation to map any one 
geodesic segment 
into a geodesic segment of $B$ real, or equivalently $\tau$ purely imaginary.
Since $SL(2, {\bf R})$ is a symmetry of the Type IIB supergravity equations, 
the most general solution is obtained by taking $B$ real, and 
then applying the most general $SL(2,{\bf R})$ transformation to the solution.
Henceforth, we restrict to $B$ real.
The reality of $B$ implies that the following quantities are real,
\bea
ac, ~ bd, ~ ad, ~ bc ~ \in ~ {\bf R}
\eea
As a result, $a/b$ and $c/d$ are real, and thus have pairwise identical
phases. The reality of the products then further imposes the following
phase arrangements,
\bea
a= a_r e^{i \theta} & \hskip 1in & c = c_r e^{-i \theta}
\no \\
b= b_r e^{i \theta} & \hskip 1in & d = d_r e^{-i \theta}
\eea
where $a_r, b_r, c_r, d_r, \theta$ are all real. It follows from (\ref{eq11})
that, unless the dilaton is constant, $c$ cannot vanish identically. 
Now substitute the above phase relations in the field equation for $c$ and derive 
that $\theta '=0$. Hence $\theta$ is a constant phase. Changing this 
phase is equivalent to making a $U(1)_\beta$ rotation. 
Thus, we may now take  also $a,b,c,d \in {\bf R}$. The general solution will
be obtained from the real solution by making the inverse 
$U(1)_\beta \times SL(2, {\bf R})$ rotation.

\subsection{Reduced equations for real $B, ~a,~b,~c,~d$}

The field equations simplify for real field, and we have, 
\bea
\label{reqabcd}
a' 
&=& -\Big( {f_1'\over f_1}+2 {f_2'\over f_2}\Big) a + 3{f_4\over f_1} b  - f^2 B' c
\no \\
b' 
&=& -\Big( 4{f_4' \over f_4}+2{f_2'\over f_2} +{f_1'\over f_1}\Big)  b 
+3 {f_4 \over f_1} a - f^2 B' d + 4 f_4 f_5 a
\no \\
c' 
&=& - \Big( {f_1'\over f_1}+2 {f_2'\over f_2}\Big) c
- 3 {f_4\over f_1} d  - f^2 B' a
\no \\
d' 
&=&  -\Big( 4{f_4' \over f_4}+2{f_2'\over f_2}+{f_1'\over
f_1}\Big) d- 3 {f_4\over f_1} c +  4 f_4 f_5  c  - f^2 B' b
\eea
Einstein's equations are arranged as follows: (\ref{E2g}) is a linear combination of
for (\ref{E1g}), (\ref{E3g}), (\ref{E4g}) and (\ref{integral}),
and will therefore be omitted,
\bea
\label{reqE}
0 &=& {f_4 '' \over f_4} 
+ 3  + 2 {(f_4')^2 \over f_4 ^2} + {f_1' f_4' \over f_1 f_4} + 4 {f_2' f_4' \over f_2 f_4}
- 4  f_5^2 f_4^2 - \half f_4^2( a^2 + c^2 + b^2+ d ^2)
\\
0&=&{f_1 '' \over f_1 } + 3  {f_1' f_4' \over f_1 f_4}
+ 4 {f_1' f_2 ' \over f_1 f_2 }  - 4 {f_1^2 f_4^2 \over f_2^4}
+ 4 f_5^2 f_4^2 + \half f_4^2( 3 a^2 -b^2 + 3 c^2- d^2) 
\nonumber \\
0&=& {f_2'' \over f_2 } -{ 6 f_4^2 \over f_2^2}   
+ 3 {f_2 ' f_4 ' \over f_2 f_4} + {f_1 ' f_2' \over f_1 f_2 }
+ 2 {f_1^2 f_4^2 \over f_2^4} + 3 {(f_2')^2 \over f_2 ^2 } 
+ 4 f_5^2 f_4^2  +\half f_4^2( a^2 +c^2 +b^2+d^2) 
\no
\eea
The constraint is given by
\bea
\label{reqconstraint}
0 & = &
12  {(f_2')^2 \over f_2 ^2 } + 12  {(f_4')^2 \over f_4 ^2} 
+ 8  {f_1 ' f_2' \over f_1 f_2 }  + 8  {f_1' f_4' \over f_1 f_4}  
+ 32 {f_2' f_4' \over f_2 f_4} - 2 f^4 (B')^2  \nonumber \\ 
&&+12 + 4  {f_1^2 f_4^2 \over f_2^4} -24 { f_4^2 \over f_2^2} +8 f_4^2 f_5^2 + 2f_4^2(  a^2 - b^2 + c^2  -d^2)
\eea
The Bianchi identity for $F_{(5)}$ is solved in terms of a single constant $C_1$,
\bea
\label{f5}
f_5 = {1 \over 6} f_1 (a^2 - c^2) + {3 \over 2} {C_1 \over f_1 f_2 ^4}
\eea
The dilatino equation simplifies as follows,
\bea
\label{req12}
c ^2 - d^2 & = & f^4 (B')^2 (f_4)^{-2} 
\eea
The remaining supersymmetry variation equations continue to be given by
(\ref{eq21}), (\ref{eq22}), (\ref{eq31}), (\ref{eq32}), and (\ref{eq4}), but
now for $a,b,c,d$ and $B$ real.

\newpage

%%%%%%%%%%%%%%%%%%%%%%%%%%%%%%%%%%%%%%%%%%
%%%%%%%%%%%%%%%%%%%%%%%%%%%%%%%%%%%%%%%%%%
\section{Solving the field and supersymmetry equations}
\setcounter{equation}{0}
%%%%%%%%%%%%%%%%%%%%%%%%%%%%%%%%%%%%%%%%%%
%%%%%%%%%%%%%%%%%%%%%%%%%%%%%%%%%%%%%%%%%%

We shall now apply the following procedure to the solution of the 
susy variation equations and the field equations: (1)~Take the solution
of the $B$-equation from the results above; (2)~Solve the susy 
variation equations and the Hamiltonian constraint (which will restrict
the possible initial data); (3)~Use those to solve the field equations.

\subsection{Solving for real $B, a,b,c,d $}

From the linear combinations of (\ref{eq21}), (\ref{eq22}), (\ref{eq31}), 
(\ref{eq32}) that expose the combinations $c \pm d$ we obtain the 
following equations for $a\pm b$, 
\bea
\label{abeq}
(a+b)(c+d) & = & 
2 {f^2 B' \over  f_4^2 } 
\left ( {f_2 ' \over f_2} + {f_1 f_4 \over f_2^2} - f_4 f_5 \right )
\no \\
(a-b)(c-d) & = & 
2 {f^2 B' \over  f_4^2 } 
\left ( {f_2 ' \over f_2} - {f_1 f_4 \over f_2^2} + f_4 f_5 \right )
\eea
The remaining equations may be expressed as follows,
\bea
\label{eqad}
3 {f_4 \over f_1}  - {f_1 f_4 \over f_2^2} - 2 f_4 f_5
& = & {2 f_4 ^2 \over f^2 B'} \, ad
\\
\label{eqac}
{f_1 ' \over f_1} + {f_2 ' \over f_2} 
& = & {2 f_4 ^2 \over f^2 B'} \, ac
\eea
They will be important in the next subsection.

\subsection{Solving the constraint}

From the constraint (\ref{reqconstraint}), the combination involving $a,b,c,d$ 
and $f^2B'$, is eliminated by using the dilatino equation (\ref{req12}) for 
$c,d$ and the $f^2B'$ term, and the product of the two equations
in (\ref{abeq}) for $a,b$, yielding
\bea
 3 \left ( {f_2 ' \over f_2} +  {f_4 ' \over f_4} \right )^2 
+ 2 \left ( {f_2 ' \over f_2} +  {f_4 ' \over f_4} \right ) 
	\left ( {f_1 ' \over f_1} +  {f_2 ' \over f_2} \right )
+3 -   {f_1^2 f_4^2 \over f_2^4} -6 { f_4^2 \over f_2^2}  
+ 4 {f_1 f_4^2 f_5 \over f_2^2} =0
\eea
Next, we use (\ref{eq4}) to replace the first term, 
\bea
\left ( {f_2 ' \over f_2} +  {f_4 ' \over f_4} \right ) 
	\left ( {f_1 ' \over f_1} +  {f_2 ' \over f_2} \right )
 +   {f_1^2 f_4^2 \over f_2^4} -3 { f_4^2 \over f_2^2}  
+ 2 {f_1 f_4^2 f_5 \over f_2^2} =0
\eea
Using the lhs of (\ref{eqad}), and the fact that $c \not= 0$
in the equation above, we find,
\bea
\label{factors}
a  \left \{ c \left ( {f_2 ' \over f_2} + {f_4 ' \over f_4} \right ) 
- d {f_1 f_4 \over f_2^2} \right \} =0
\eea 
Thus, the constraint factorizes when reduced to the subspace of supersymmetric 
solutions. We study the vanishing of each factor in turn. 

\subsection{The Case $a=0$}\label{secazero}

The vanishing of $a$ in (\ref{eqac}) and (\ref{eqad})
together with the $c \not= 0$ implies that,
\bea
f_1 f_2 =  \rho
\hskip 1in
f_5 = {3 \over 2 f_1} - \half {f_1 \over f_2^2}
\label{f1f2}
\eea
with $\rho$ a constant, which will be fixed later. The remaining differential equation for $f_2$ is,
\bea
\label{eqf_2'}
2 c {f_2 ' \over f_2} = 3d \left ( {f_1f_4 \over f_2^2} - {f_4 \over f_1} \right )
\eea
The reduced field equation for $a$ in (\ref{reqabcd}) for $a=0$ simplifies to,
\bea
\label{beq}
b = {f_1 \over 3 f_4 } f^2 B' \, c
\eea
Using this expression to eliminate $b$ from (\ref{eq21}) and (\ref{eq22}) yields 
$c$ and $d$, 
\bea
\label{ceq}
c & = & 3 \left ( {1 \over f_2^2} - {1 \over f_1^2} \right )^\half 
\\
\label{cdeq}
cd & = & 6 \, {f_2 ' \over f_1 f_2 f_4}
\eea
Combining the result for $c$ with the $f_5$ equation in (\ref{f5}),
we find that $f_1^2 f_2 ^2 = \rho ^2 = 3 C_1/2$.

\medskip

The field equations for $a,b,c,d$ of (\ref{reqabcd}) are now satisfied as follows.
The equation for $a$ is automatic using (\ref{beq}); the equation for $b$ is 
automatic using (\ref{beq}); the equation for  $c$ is automatic using (\ref{cdeq}).
Finally, the equation for $d$ is handled as follows. Start with the dilatino equation
(\ref{req12}) as a definition of $d$, take the derivative and use the field equations
in (\ref{reqabcd}) for $c, d$. One arrives at
\bea
\label{cdf2}
4 (c^2 - d^2) {f_2 ' \over f_2} - 8 c^2 {f_4 ' \over f_4} + 8 f_4 f_5 cd - 2 f^2 B' bd
=0
\eea
Using (\ref{beq}) to eliminate $b$ implies that the first and last
terms in (\ref{cdf2}) cancel one another, leaving the following linear 
relation between  $c$ and $d$, after eliminating $f_5$ using (\ref{f1f2}),
\bea
c {f_4 ' \over f_4} 
= \left ( {3 \over 2} {f_4 \over f_1} - \half {f_1 f_4 \over f_2^2} \right ) d
\eea
Eliminating now $f_4/f_1$ using (\ref{eqf_2'}), we get
\bea
\label{cdrelation}
c \left ( {f_2 ' \over f_2} + {f_4 ' \over f_4} \right ) 
- d {f_1 f_4 \over f_2^2} =0
\eea
This is the second factor in the relation (\ref{factors}). 

\subsection{Exact solution of the case $a=0$ via hyper-elliptic integrals}

Combining the relations (\ref{cdrelation}) and  (\ref{req12}), 
we may solve completely for $c$ and $d$,
\bea
c ^2 =  {C_2^2 \over f_1^4 f_2^8 f_4^6}
\hskip 1in 
d^2 = 
{C_2^2 \over f_1^4 f_2^8 f_4^6} \left ( 1 - {f_2 ^4 \over f_1^2 f_4^2}
\right ) \label{cdeqpsi}
\eea
Eliminating $c$ between the expression above and the one already obtained in (\ref{ceq}), we obtain a polynomial relation between $f_1$, $f_2$, and $f_4$,
\bea
\label{algrel}
\left ( f_1 ^4 f_2^6 - f_1 ^2 f_2^8 \right ) f_4 ^6 = {1 \over 9} C_2^2
\eea
The function $f_1$ may be eliminated using (\ref{f1f2}), and the resulting 
equations (\ref{eq4}) and (\ref{algrel}) may be expressed in terms of 
$f_4$ and the composite $\psi$  defined by $f_2 f_4 = \rho /\psi$, so that
\bea
f_4 ^4 & = & {\rho^2 \over \psi ^4 } + {C_2^2 \over 9 \rho ^6} \psi ^2
\label{ffourpsi}\\
\left ( { \psi ' \over \psi } \right )^2 
& = & \rho^{-4}  f_4^8 \psi ^6 -1
\eea
Eliminating $f_4$ gives a single genus 5 decoupled hyper-elliptic equation for $\psi$, 
or equivalently gives a genus 3 decoupled equation for $\Psi \equiv \psi ^2$, 
\bea
(\psi ')^2 & = &  \left ( 1 + {C_2^2 \over 9 \rho ^8}  \psi ^6 \right )^2 -\psi ^2 
\label{psieqb} \\
{1 \over 4} (\Psi ')^2 & = &
\Psi \left ( 1 + {C_2^2 \over 9 \rho ^8}  \Psi ^3 \right )^2 -\Psi ^2 
\label{psieqa}
\eea

\subsection{The general Case}

The general case is specified by the vanishing of the second factor in 
(\ref{factors}) only. This readily allows for the solution of $c$ and $d$
in terms of the functions $f_1, f_2, f_4$, and we find,
\bea
c = { C_2 \over f_1 ^2 f_2 ^4 f_4^3}  
\hskip 1in 
d = v c 
\hskip 1in  v = \left (1 - {f_2 ^4 \over f_1 ^2 f_4 ^2}  \right )^\half
\eea
The remaining susy variation equations are equivalent to
\bea
2 a \, {f_1 f_4^2 \over f_2^2} = 
+{f_1 ' \over f_1}  + {f_2 ' \over f_2} ~
& \hskip .8in &
2 a v \, {f_1 f_4^2 \over f_2^2} = 
+3 {f_4 \over f_1} - {f_1 f_4 \over f_2^2} - 2 f_4 f_5
\no \\
2 bv  \, {f_1 f_4^2 \over f_2^2} = 
-{f_1 ' \over f_1}  + 3 {f_2 ' \over f_2} 
& \hskip .8in &
~ 2 b \, {f_1 f_4^2 \over f_2^2} = 
-3 {f_4 \over f_1} +5 {f_1 f_4 \over f_2^2} - 2 f_4 f_5
\eea
Eliminating $a,b$ gives
\bea
v \left ( +{f_1 ' \over f_1}  + {f_2 ' \over f_2}  \right ) & = & 
+3 {f_4 \over f_1} - {f_1 f_4 \over f_2^2} - 2 f_4 f_5
\no \\
{1 \over v} \left ( -{f_1 ' \over f_1}  + 3 {f_2 ' \over f_2}  \right ) & = & 
-3 {f_4 \over f_1} +5 {f_1 f_4 \over f_2^2} - 2 f_4 f_5
\eea
Finally, eliminating $f_5$ and $a$ using (\ref{f5})  gives the following equations,
\bea
\left ( v + {1 \over v} \right ) {f_1 ' \over f_1}
+ \left ( v - {3 \over v} \right ) {f_2 ' \over f_2}
& = &
6 \left ( {f_4 \over f_1} - {f_1 f_4 \over f_2^2} \right )
\no \\
\left ( 
{f_1 ' \over f_1} + {f_2 ' \over f_2} + 6 {f_1 f_4^3 \over f_2^4}v \right )^2
& = &
36 {f_1 ^2 f_4^6 \over f_2^8} - 12 {f_1 ^2 f_4^4 \over f_2^6}
+ 4 { C_2 ^2 \over f_1^2 f_2^{12} f_4^2} - 36 { C_1 f_4^4 \over f_2^8}
\eea
In attempting to separate these three equations, it appears natural to
define the following new combinations  of the functions $f_1, f_2, f_3$, 
\bea
\psi _1 \equiv f_2 f_4 
\hskip .7in 
\psi _2  \equiv f_1 f_2 
\hskip .7in 
\psi _3 \equiv {f_1 f_4 \over f_2^2}  
\hskip .7in 
v= \sqrt{1 - \psi _3 ^{-2}}
\eea
Together with equation (\ref{eq4}), the equations then become,
\bea
{\psi _1 ' \over \psi _1 } 
& = & 
\psi _3 v
\no \\
v {\psi _2 ' \over \psi _2} + {1 \over v} {\psi _3 ' \over \psi _3} 
& = & 
-5 \psi _3 + 6 {\psi _1 \over \psi _2}
\no \\
\left ( {\psi _2' \over \psi _2} + 6 {\psi _1 \psi _3 ^2 \over \psi _2} v \right )^2 
& = &
36 {\psi _1 ^2 \psi _3 ^4 \over \psi _2 ^2} 
- 12 {\psi _1  \psi _3 ^3 \over \psi _2} 
+ 4 {C_2^2 \psi _3 ^2 \over \psi _1 ^4 \psi _2^4} 
- 36 C_1 {\psi _1 \psi _3 ^3 \over \psi _2^3}
\eea
So far, we have not succeeded in decoupling these equations.

\newpage

%%%%%%%%%%%%%%%%%%%%%%%%%%%%%%%%%%%%%%%%%%
%%%%%%%%%%%%%%%%%%%%%%%%%%%%%%%%%%%%%%%%%%
\section{Numerical results}
\setcounter{equation}{0}
For the  special case $a=0$  it was shown in section \ref{secazero} that the
existence of an unbroken supersymmetry is equivalent to the existence
of solutions to the equation (\ref{psieqb}) which can be viewed as describing
the motion if a particle with coordinate $\psi$ in a potential $V(\psi)$.
\be
(\psi')^2+ V(\psi)=0
\ee 
where the potential is given by
\be
V(\psi) = -\Big(1+{C_2^2\over  9 \rho^8}\psi^6\Big)^2+\psi^2
\ee

\begin{figure}[tbph]
\begin{center}
\epsfxsize=2.4in
\epsfysize=2.0in
\epsffile{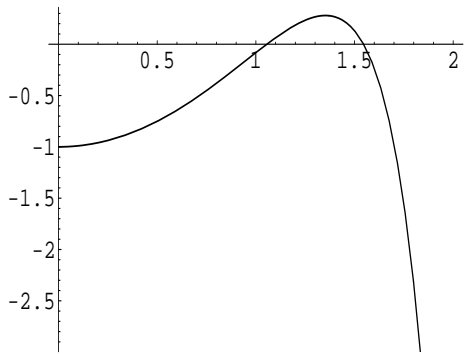}
\label{fig3}
\caption{The potential $V(\psi)$ for $C_2^2/\rho^8 =.36$}
\end{center}
\end{figure}

\begin{figure}[tbph]
\begin{center}
\epsfxsize=5.0in
\epsfysize=2.1in
\epsffile{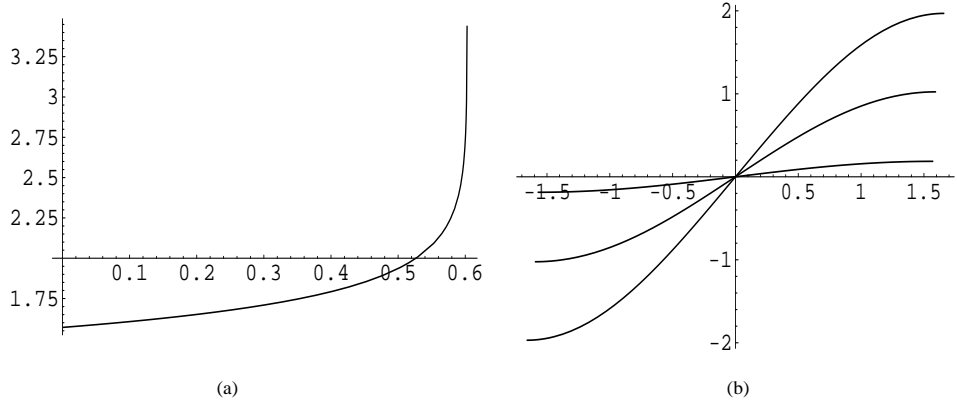}
\caption{(a) Value of $\mu_0$ as a function of $C_2^2/\rho^8$, (b)
dilaton  for three values of parameters}
\label{figb}
\end{center}
\end{figure}

The complete form of the solution is determined once $\psi$ is found.  
Near the  boundary of AdS the metric function  $f_4$ behaves like $f_4  \sim
1/(\mu \mp \mu_0)$ as $\mu\to \pm \mu_0$. It follows from (\ref{ffourpsi}) that
$\psi$ vanishes as $\psi \sim
(\mu\mp \mu_0)$ in this limit. On the other hand the limit $\psi\to
\infty$ corresponds to the metric  becoming singular. A regular
Janus like solution can only exist  if the potential is
somewhere positive, since in this case there are two allowed regions $0<\psi<
\psi_1$ and $\psi_2< \psi< \infty$. For the   nonsingular Janus solution 
$\psi$  takes values in the first region.
It is easy to see that this implies a condition
on the parameters of the potential
\be
{C_2^2\over \rho^8}< {5^5 \over  2^6 3^4} \label{critrange}
\ee
The metric for the undeformed $AdS_5$ corresponds to
$f_4(\mu)=1/\cos(\mu)$ with range $\mu \in [-\pi/2,\pi/2]$. For the Janus
solution the range $\mu$ is increasing  
$\mu\in [-\mu_0,\mu_0]$,  where $\mu_0 > \pi/2$. The range depends on
$C_2^2/\rho^8$ and diverges when this parameter approaches its critical
value (\ref{critrange}) (See Figure \ref{figb} (a)). 

\smallskip

In the following we present the results for the solution for three choices of 
parameters $\rho$ and $C_2$. We have fixed  $\rho=1.4$ and picked 
$C_2=0.445, C_2= 0.245$ and $C_2=0.045$.
The dilaton can be determined by  numerically integrating (\ref{dileqa}), the
plot of the dilaton for the three choices of parameters is given in 
Figure \ref{figb} (b),

The metric function $f_4$ is related to $\psi$ by equation (\ref{ffourpsi}), it
diverges at $\mu =\pm \mu_0$ which corresponds to the two AdS boundary
components.  The metric function $f_2$ and $f_1$ are determined by the
relations $f_2 f_4=\rho/\psi$ and $f_1 f_2=\rho$.

\begin{figure}[tbph]
\begin{center}
\epsfxsize=5.0in
\epsfysize=2.0in
\epsffile{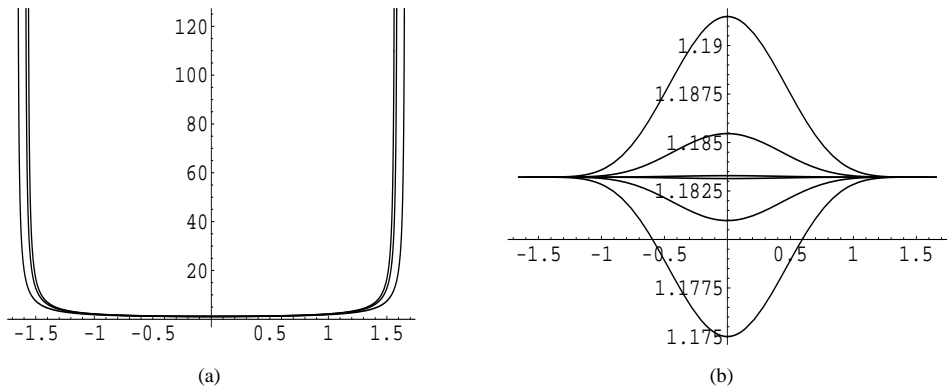}
\caption{(a) $f_4$ for three values of parameters, (b) $f_1$ and
$f_2$ for three values of parameters }
\label{figure5}
\end{center}
\end{figure}
Note that since $f_1$ is the scale factor of the $S^1$ fiber and
$f_2$ is the scale factor of the $CP_2$ base in the squashed sphere,
the fact that $f_1\to f_2$ as one approaches the AdS boundary components  $\mu\to \pm \mu_0$, means that the
sphere becomes 'un-squashed' and the $SO(6)$ symmetry is restored at
the AdS boundary.
The third rank anti-symmetric tensor is determined by $\psi$ via
equation (\ref{cdeqpsi}) and the plot for the two function $c$ and $d$  is given by

\begin{figure}[tbph]
\begin{center}
\epsfxsize=5.0in
\epsfysize=2.0in
\epsffile{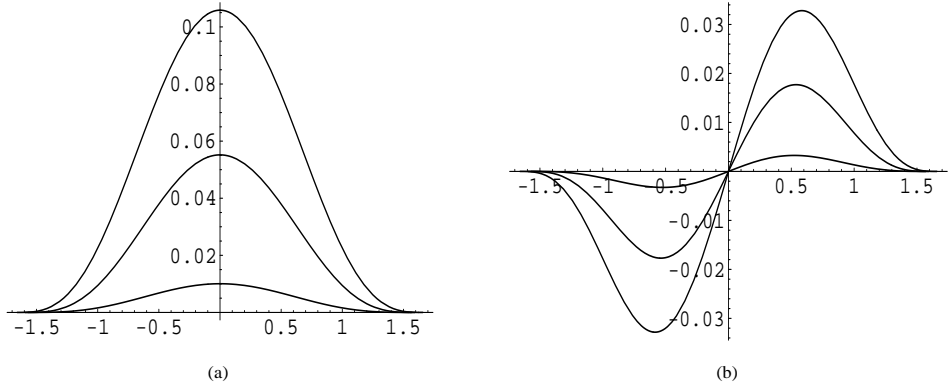}
\label{figure6}
\caption{(a) AST function $c$  for three values of parameters, (b) AST function $d$  for three values of parameters}
\end{center}
\end{figure}

\newpage

%%%%%%%%%%%%%%%%%%%%%%%%%%%%%%%%%%%%%%%%%%
%%%%%%%%%%%%%%%%%%%%%%%%%%%%%%%%%%%%%%%%%%
\section{Holographic dual}
\setcounter{equation}{0}
%%%%%%%%%%%%%%%%%%%%%%%%%%%%%%%%%%%%%%%%%%
%%%%%%%%%%%%%%%%%%%%%%%%%%%%%%%%%%%%%%%%%%

The AdS/CFT correspondence relates 10-dimensional Type IIB supergravity 
fields to gauge invariant operators on the $\N=4$ super Yang-Mills side. In the
following we  briefly review some aspects of this map. The Poincar\'e metric of Euclidean $AdS_5$ is given by
\be
ds^2= {1\over z^2} \left (dz^2 + \sum_i dx_i^2 \right )
\ee
Near the boundary of $AdS_5$, where $z\to 0$, 
a scalar field $\Phi_m$ of mass $m$  behaves as
\be
\Phi _m (z,x) \, \sim \, \phi_{non-norm}(x) z^{4-\Delta} +  \phi_{norm}(x) z^{\Delta}  
\label{stateoper}
\ee
where $m^2=\Delta(\Delta-4)$. The non-normalizable mode corresponds to
insertion in the Lagrangian of an operator ${\cal O}_\Delta$ with scaling 
dimension $\Delta$.
The boundary source can be determined from (\ref{stateoper}) by
\be
\phi_{non-norm}(x) = \lim_{z\to 0}   z^{\Delta-4} \Phi(z,x) \label{nonnorma}
\ee
If $\phi_{non-norm}$ vanishes a nonzero $\phi_{norm}$ corresponds to a
non vanishing expectation value $\langle{\cal O}_\Delta\rangle= \phi_{norm}$ of the
operators ${\cal O}_\Delta$ on the Yang-Mills side.

\smallskip

As reviewed in section (\ref{secadscftdual}) for the Janus solution,
the boundary geometry, and hence the holographic dual, are more complicated. 
The asymptotic behavior of the solution obtained in section (\ref{secazero}) is 
readily obtained using power series expansion near the boundary components.

\smallskip

Employing Poincar\'e coordinates for the $AdS_4$ slices, the asymptotic 
form of the non compact part of the ten dimensional metric can be obtained by expanding   $f_4(\mu)$ near the two boundary components $\mu=\pm\mu_0$
\be
ds^2= {1\over (\mu\mp \mu_0)^2 z^2 } \left ( dz^2 + \sum_{i=1}^3 dx_i^2 + z^2 d
\mu^2 \right )+  O [(\mu\mp \mu_0)^0] 
\label{asymmetric}
\ee
The boundary is reached when $(\mu\mp \mu_0) z \to 0$, and
consists of one component  with $\mu = \mu_0$ and one component with 
$\mu= - \mu_0$. The complete boundary  corresponds to two
4-dimensional half spaces joined by a ${\bf R}^3$ interface located at $z=0$. 

\smallskip

The asymptotic behavior of the dilaton  near the boundaries is given by
\be
\phi(\mu)= \phi_0^{\pm} + {C_2 \over 2 \rho^4}(\mu\mp \mu_0)^4 
+O [(\mu\mp\mu_0)^6]
\label{asymdil}
\ee
The dilaton corresponds to a dimension $\Delta=4$ operator. Hence it follows
from (\ref{nonnorma}) that there is a source $\phi_0^\pm$ of the operator dual
to the dilaton in the two boundary components.
This is interpreted on the Yang-Mills side as a theory where  the gauge coupling takes two different values on the half spaces separated by a planar interface.

\smallskip

The asymptotic behavior of the anti-symmetric tensor field components 
$c$ and $d$ is,  
\bea
c(\mu)&= & {C_2\over \rho^{9/ 2}} (\mu_0 \mp \mu)^3 +O[(\mu\mp\mu_0)^5]
\no \\
d(\mu)&= & {C_2\over \rho^{9/ 2}} ( \mu\mp \mu_0)^3 +O[(\mu\mp\mu_0)^5]
\label{asymc}
\eea
The $(\mu_0 \mp \mu)^3$ behavior of the  functions $c$ and $d$ is in
agreement with the fact that the lowest Kaluza-Klein modes on the $S^5$ of the
anti-symmetric rank 2 tensor field  is associated with dimension $\Delta=3$
operator \cite{Kim:1985ez,Gunaydin:1984fk}. 

\smallskip

In the light of (\ref{nonnorma}),  the $c\sim (\mu\mp \mu_0)^\Delta$ behavior
of (\ref{asymc}) seems to suggest that there is no source for the dual
operator of the anti-symmetric tensor. 
However this conclusion is  not correct. For the Janus metric the
appropriate rescaling of the field to extract the non normalizable mode is
given by
\bea
c_{non-norm}&=& \lim_{\epsilon\to 0} \epsilon^{\Delta-4} c(\mu) \no \\
&=& \lim_{\epsilon \to  0} {1\over \epsilon }  {C_2\over
  \rho^{9/2}} (\mu \mp \mu_0)^3  \no \\
&=& \lim_{(\mu\mp \mu_0) z\to 0}   {\ (\mu\mp \mu_0)^2 \over z} {C_2\over
  \rho^{9/2}} \label{cnonnorm}
\eea
where $\epsilon= (\mu\mp \mu_0) z$ was used.
For a point on the boundary which is away from the three dimensional interface
one has $z\neq 0$ and it follows from (\ref{cnonnorm}) that the  source
for the dual operator vanishes away from the interface.
However for the interface one has $z=0$ and $c_{non-norm}$ in
(\ref{cnonnorm}) diverges. This behavior indicates  the presence of a
delta function source for the dual $\Delta=3$ operator on the interface, since 
the integral over a small disk around the interface $ \int d\mu\; dz z\;
c_{non-norm}$  is finite. 

\smallskip

The metric functions $f_1$ and $f_2$ behave as
\bea
f_1(\mu) &= & \sqrt{\rho}\left (
1+ {C_2^2 \over 36 \rho^{8}} (\mu\mp\mu_0)^6+ O[(\mu\mp\mu_0) ^8]  \right ), 
\no \\
f_2 (\mu) &= & \sqrt{\rho}\left (
1-{C_2^2 \over 36 \rho^{8}}(\mu\mp\mu_0)^6+ O[(\mu\mp\mu_0)^8] \right )
\eea 
Repeating the analysis of the anti-symmetric tensor for the metric fields
reveals that there no 
additional operator sources turned on by $f_1$ and $f_2$.

%%%%%%%%%%%%%%%%%%%%%%%%%%%%%%%%%%%%%%%%%%
%%%%%%%%%%%%%%%%%%%%%%%%%%%%%%%%%%%%%%%%%%

%%%%%%%%%%%%%%%%%%%%%%%%%%%%%%%%%%%%%%%%%%
%%%%%%%%%%%%%%%%%%%%%%%%%%%%%%%%%%%%%%%%%%
\section{Conclusions}
\setcounter{equation}{0}

In this paper, a supersymmetric generalization of the Janus solution of
ten dimensional type IIB superstring theory was found. The solution
was constructed by imposing the condition of the existence of a preserved
supersymmetry on the most general ansatz compatible with
$SO(2,3)\times SU(3)\times U(1)_\beta \times SL(2, {\bf R})$ symmetry. 
The supergravity solution is in agreement  with the field theoretical analysis of
\cite{Clark:2004sb,edjemg}. The restoration of supersymmetry requires a 
nontrivial antisymmetric tensor $B_{(2)}$.  The dual holographic interpretation 
of the presence of the $B_{(2)}$ tensor field is given by turning on a dimension $\Delta=3$ operator on the interface.  There are several questions relating to 
our work which are worth pursuing:

The generalized non-supersymmetric Janus solution we have found can be represented exactly using elliptic functions. Therefore, it may be possible to 
calculate correlation functions in our background exactly, and compare supergravity and field theory predictions beyond conformal perturbation theory.

In \cite{edjemg} a detailed analysis of the interface field theory found that
there are additional possibilities to obtain a larger unbroken supersymmetry.
In particular a case with $\N=4$ interface supersymmetry was found. The
internal symmetry of this theory is $SU(2)\times SU(2)$. A possible Ansatz in
this case replaces the squashed five sphere used in this paper with the
$T^{1,1}$ manifold \cite{Romans:1984an,Klebanov:1998hh} which realizes the
$SU(2)\times SU(2)$ symmetry. 

The $AdS_5\times S_5$ background is famously obtained as the near
horizon limit of $N$ D3-branes. Various defect conformal field
theories can holographically be obtained via  near horizon limits of
intersecting D-brane systems. The brane interpretation of the
non-supersymmetric Janus, as well as our supersymmetric 
solution, is unknown at this point. The fact that both the dilaton as
well as the AST field are sourced by fivebranes suggests that a
 brane realization of the supersymmetric Janus solution  could be 
given by the  near horizon limit of intersecting D3/D5 branes.
However, the only known solutions of this kind treat the five branes in
the probe approximation \cite{Karch:2000gx,DeWolfe:2001pq}. Whether
a fully back reacted solution is related to  Janus is a very interesting
question, which we plan to investigate in the future.

%%%%%%%%%%%%%%%%%%%%%%%%%%%%%%%%%%%%%%%%%%
%%%%%%%%%%%%%%%%%%%%%%%%%%%%%%%%%%%%%%%%%%

\vskip .3in

\noindent{\Large \bf Acknowledgments}

\vskip .1in

\noindent 
It is a pleasure to acknowledge helpful conversations with Iosif Bena, Per 
Kraus, and Norisuke Sakai. This work was supported in part by National 
Science Foundation (NSF) grant PHY-04-56200.
ED is grateful to the Kavli Institute for Theoretical Physics (KITP)
for their hospitality and support under NSF grant PHY-99-07949.
MG is grateful to the Harvard Particle Theory group for hospitality
while this work was being completed.

\newpage

\appendix

\section{Realization of the $\Gamma$-matrices}

It will be useful to choose a well-adapted basis for $\G$-matrices,
\bea
\G ^\mu & = & \sigma _1 \otimes \gamma ^\mu \otimes I_4 \hskip 1in \mu = 0,1,2,3,4
\no \\
\G ^ m & = & \sigma _2 \otimes I_4 \otimes \gamma ^m \hskip 1in m=5,6,7,8,9
\eea
for which we have 
\bea
\label{chiralBapp}
\G^{11} & = &  \G^{0123456789} = \sigma _3 \otimes I_4 \otimes I_2 \otimes I_2  
\no \\
\B & = & \G^{2568} \hskip .35in = \sigma _3 \otimes \gamma ^2 \otimes \sigma _1 \otimes \sigma _2
\eea
The following convention for the $4 \times 4$ matrices $\gamma ^A$ will be adopted,
\bea
i\gamma ^0 =  \sigma _2 \otimes I_2
& \hskip 1in & 
\gamma ^5 = \sigma _3 \otimes \sigma _3
\no \\
\gamma ^1 = \sigma _1 \otimes I_2
& \hskip 1in &
\gamma ^6 = \sigma _1 \otimes I_2 
\no \\
\gamma ^2 = \sigma _3 \otimes \sigma _2
& \hskip 1in &
\gamma ^7 = \sigma _2 \otimes I_2
\no \\
\gamma ^3 = \sigma _3 \otimes \sigma _1
& \hskip 1in &
\gamma ^8 = \sigma _3 \otimes \sigma _1
\no \\
\gamma ^4 = \sigma _3 \otimes \sigma _3
& \hskip 1in & 
\gamma ^9 = \sigma _3 \otimes \sigma _2
\eea
The $\G$-matrices in the complex frame 
associated with $CP_2$ and (\ref{complexframe}), are as follows,
\bea
\G^{z_1} = \G^6 + i \G^7 & = & 
2 \sigma _2 \otimes I_4 \otimes \sigma _+ \otimes I_2
\no \\
\G^{\bar z_1} = \G^6 - i \G^7 & = & 
2 \sigma _2 \otimes I_4 \otimes \sigma _- \otimes I_2
\no \\
\G^{z_2} = \G^8 + i \G^9 & = & 
2 \sigma _2 \otimes I_4 \otimes \sigma _3 \otimes \sigma _+
\no \\
\G^{\bar z_2} = \G^8 - i \G^9 & = &
2 \sigma _2 \otimes I_4 \otimes \sigma _3 \otimes \sigma _-
\eea
The following combinations will also be useful in evaluating the connection form,
\bea
\G^{ij} & = & I_2 \otimes \gamma ^{ij} \otimes I_4
\hskip 1in i,j =0,1,2,3
\no \\
\G^{i4} & = & I_2 \otimes \gamma ^i \gamma ^4 \otimes I_4
\no \\
\G ^{45} & = & i \sigma _3 \otimes \gamma ^4 \otimes \gamma ^5
\no \\
\G^{4a} & = & i \sigma _3 \otimes \gamma ^4 \otimes \gamma ^a
\no \\
\G^{5a} & = & I_2 \otimes I_4 \otimes \gamma ^5 \gamma ^a
\no \\
\G^{ab} & = & I_2 \otimes I_4 \otimes \gamma ^{ab}
\hskip 1in a,b=6,7,8,9
\eea

\end{document}